\documentclass[utf8]{frontiersSCNS} % for Science, Engineering and Humanities and Social Sciences articles
\usepackage{url,hyperref,lineno,microtype,subcaption,epstopdf,aas_macros}
\usepackage[onehalfspacing]{setspace}
\usepackage{verbatim}
\usepackage{bm}		          % allow shorthand to bold text
\newcommand\thetanz{$\theta = 90^\circ$}
\newcommand\thetaz{$\theta = 0^\circ$}
\newcommand\sm{$\sim$}
\newcommand\Htwo{$\sim$}
\newcommand\citepeg[1]{\citep[e.g.][]{#1}}
\newcommand\Msun{M$_\odot$} 
\newcommand{\alfven}{Alfv{\'e}n}
\newcommand\cms{cm~s$^{-1}$} 
\newcommand{\phantomsph}{\textsc{Phantom}}
\newcommand\zetaeq[1]{$\zeta_\text{cr} = 10^{#1}$ s$^{-1}$}  
\newcommand\zetacr{$\zeta_\text{cr}$}

\def\keyFont{\fontsize{8}{11}\helveticabold }
\def\firstAuthorLast{Wurster \& Li} %use et al only if is more than 1 author
\def\Authors{James Wurster\,$^{1}$ and Zhi-Yun Li\,$^{2}$}
\def\Address{$^{1}$School of Physics and Astronomy, University of Exeter, Stocker Rd, Exeter EX4 4QL, UK \\
$^{2}$Astronomy Department, University of Virginia, Charlottesville, USA }
\def\corrAuthor{James Wurster}

\def\corrEmail{j.wurster@exeter.ac.uk}

\begin{document}
\onecolumn
\firstpage{1}
\title[Magnetic fields in protostellar discs]{The role of magnetic fields in the formation of protostellar discs\footnote{Published, open access version available at \href{https://www.frontiersin.org/articles/10.3389/fspas.2018.00039/full}{www.frontiersin.org/articles/10.3389/fspas.2018.00039/full}.}}

\author[\firstAuthorLast ]{\Authors} %This field will be automatically populated
\address{} %This field will be automatically populated
\correspondance{} %This field will be automatically populated

\extraAuth{}% If there are more than 1 corresponding author, comment this line and uncomment the next one.
%\extraAuth{corresponding Author2 \\ Laboratory X2, Institute X2, Department X2, Organization X2, Street X2, City X2 , State XX2 (only USA, Canada and Australia), Zip Code2, X2 Country X2, email2@uni2.edu}

\maketitle

\begin{abstract}
\section{}
%%%%\texttt{NOTE: As per the website instructions, the abstract must be limited to 350 words; this is that version: Word count: 350}
The formation of a protostellar disc is a natural outcome during the star formation process.  As gas in a molecular cloud core collapses under self-gravity, the angular momentum of the gas will slow its collapse on small scales and promote the formation of a protostellar disc.  Although the angular momenta of dense star-forming cores remain to be fully characterized observationally, existing data indicates that typical cores have enough angular momenta to form relatively large, 100 au-scale, rotationally supported discs, as illustrated by hydrodynamic simulations. However, the molecular clouds are observed to be permeated by magnetic fields, which can in principle strongly affect the evolution of angular momentum during the core collapse through magnetic braking. Indeed, in the ideal magnetohydrodynamic (MHD) limit, magnetic braking has been shown to be so efficient as to remove essentially all of the angular momentum of the material close to the forming star such that disc formation is suppressed.  This failure to produce discs in idealized cores is known as the magnetic braking catastrophe.  The catastrophe must be averted in order for the all-important rotationally supported discs to appear, but when and how this happens remains debated. We review the resolutions proposed to date, with emphasis on misalignment,  turbulence and especially non-ideal effects. Non-ideal MHD accounts for charged and neutral species, making it a natural extension to the ideal MHD approximation, since molecular clouds are only weakly ionized.  The dissipative non-ideal effects diffuse the magnetic field to weaken it, and the dispersive term redirects the magnetic field to promote or hinder disc formation, dependent upon the magnetic geometry.  When self-consistently applying non-ideal processes, rotationally supported discs of at least tens of au form, thus preventing the magnetic braking catastrophe.  The non-ideal processes are sensitive to the magnetic field strength, cosmic ray ionization rate, and gas and dust grain properties, thus a complete understanding of the host molecular cloud is required.  Therefore, the properties of the host molecular cloud -- and especially its magnetic field -- cannot be ignored when numerically modelling the formation and evolution of protostellar discs.

\tiny
 \keyFont{ \section{Keywords:} magnetic fields, magnetohydrodynamics (MHD), non-ideal MHD, star formation, protostellar discs} 
 \end{abstract}

%------------------------------------------------------------------------------------------------------------------------------------------------------------------------------------------------------------------------------------------------
\section{Introduction}
\label{sec:intro}

The broad outline of low-mass star formation has been known since at least \citet{Larson1969}, although many specific details are still under investigation.  In Larson's description, which is the foundation for all current low-mass star formation models, a piece of the interstellar cloud (a molecular cloud core in modern terminology) collapses under self-gravity.  The collapse is initially isothermal, since radiation is efficiently radiated away.  However, as the density increases at the centre of the core, it becomes optically thick to the radiation, which leads to an increase in thermal pressure support against self-gravity and the formation of the first hydrostatic or first Larson core.  The first hydrostatic core continues to accrete material from the collapsing envelope, and its mass, density and temperature increase until the temperature rises above \sm2000~K; this temperature triggers the dissociation of \Htwo, allowing the core to further collapse.  This second collapse phase is rapid, and lasts until most of the \Htwo \ has been dissociated, at which point the second hydrostatic or stellar core has formed.  The temperature continues to rise until nuclear burning starts and the star is formed.

The formation of star-forming molecular cloud cores is not fully understood\footnote{The focus of this review is on disc formation, thus for the remainder of this paper, we will assume that a slowly rotating cloud core has successfully formed.}.  These cores are observed to be initially slowly rotating, with ratios of rotational energy to gravitational potential being $\beta \lesssim 0.15$ with typical values of $\beta \sim 0.02$ \citep{Goodman+1993}.  However, their angular velocities are typically one to two orders of magnitude smaller than inferred by conservation of angular momentum \citep[for a review, see][]{GoldsmithArquilla1985}.  Therefore, there must exist some mechanism that will shed the angular momentum to allow these slowly rotating cloud cores to form \citepeg{Spitzer1968}.

As the rotating cloud core collapses under self-gravity, in the absence of magnetic fields, the rotation slows the collapse such that the gas forms a large protostellar disc as early as during the first core stage, and certainly by the Class 0 phase, as indicated by observations \cite[e.g.][]{Tobin+2012,Murillo+2013,Codella+2014,Lee+2017} and found in numerical simulations \cite[e.g.][]{Boss1993,YorkeBodenheimerLaughlin1993,YorkeBodenheimerLaughlin1995,BossMyhill1995,BateTriccoPrice2014,Tomida2014,WursterBatePrice2018hd}.  Conservation laws and observations thus both suggest that protostellar discs are a natural byproduct of the star formation process.  

Molecular clouds are observed to be strongly magnetized \citep[e.g.][]{Crutcher1999,Bourke+2001,HeilesCrutcher2005,TrolandCrutcher2008}, and magnetic fields are efficient at transporting angular momentum away from a collapsing core \citep[known as `magnetic braking'; e.g.][]{MestelSpitzer1956,MouschoviasPaleologou1979,MouschoviasPaleologou1980,BasuMouschovias1994,BasuMouschovias1995b,MellonLi2008}.  On the cloud scale, magnetic braking likely occurs early in the cloud's formation and is responsible (at least in part) for reducing the angular momentum to the observed values \citepeg{Mouschovias1983}.  Near the centre of the collapsing core, magnetic braking means that discs are less necessary to conserve angular momentum since it is transported away.  This reduced angular momentum may delay the formation of the disc until during or after the stellar core phase, or may prevent it altogether.  In idealized numerical simulations including ideal magnetohydrodynamics (MHD), protostellar discs either fail to form or are much smaller than the observed sizes.  This is known as the magnetic braking catastrophe \citep{AllenLiShu2003,Galli+2006}.

Magnetic fields support charged gas against gravitational collapse, thus a common characterization of the relative importance of the gravitational and magnetic forces is the normalized mass-to-flux ratio,
\begin{flalign}
\label{eq:masstofluxmu}
\mu &\equiv \frac{M/\Phi_\text{B}}{\left(M/\Phi_\text{B}\right)_\text{crit}},&
\end{flalign}
where
\begin{flalign}
\label{eq:masstoflux}
\frac{M}{\Phi_\text{B}} &\equiv \frac{M}{\pi R^2 B},&
\end{flalign}
is the mass-to-flux ratio and 
\begin{flalign}
\label{eq:masstofluxcrit}
\left(\frac{M}{\Phi_\text{B}}\right)_\text{crit} &= \frac{c_1}{3\pi}\sqrt{ \frac{5}{G} },&
\end{flalign}
is the critical value where the gravitational and magnetic forces balance; in these equations, $M$ is the total mass contained within a core of radius $R$, $\Phi_\text{B}$ is the magnetic flux threading the surface of the spherical core assuming a uniform magnetic field of strength $B$, $G$ is the gravitational constant and $c_1 \simeq 0.53$ is a dimensionless coefficient numerically determined by \citet{MouschoviasSpitzer1976}.   The critical value of $\mu = 1$ suggests that the gravitational and magnetic forces balance one another.  For large super-critical values ($\mu \gtrsim 20$), the magnetic field is inconsequential for core collapse, and the evolution is similar to that of a purely hydrodynamic cloud \citep[e.g.][]{BateTriccoPrice2014}.  For sub-critical values ($\mu < 1$), the magnetic field will prevent the collapse of the cloud core altogether.  Observations suggest $\mu \sim 2-10$ in molecular cloud cores \citep[e.g.][]{Crutcher1999,Bourke+2001,HeilesCrutcher2005}, however, this value could be even smaller after correcting for projection effects \citep{LiFangHenningKainulainen2013}.

Although widely used, the mass-to-flux ratio should be used with caution, since the equation and the critical value are dependent on the geometry.  While the above equations assume spherical geometry, a mass-to-flux ratio for a thin sheet is given in \citet{NakanoNakamura1978}, and the ratio for an oblate spheroid is given in \citet{MouschoviasSpitzer1976}. 

%V2: Although widely used, the mass-to-flux ratio should be used with caution, since its derivation assumes a uniform magnetic field and uniform density \citepeg{Mouschovias1976a,Mouschovias1976b}.  Thus, in numerical simulations, clouds initialized with centrally condensed mass profiles (e.g. a Bonnor-Ebert sphere; \citealp{Bonnor1956,Ebert1955}) will not give a true initial value, nor will calculations at later times when density structures form and the magnetic field becomes non-uniform and tangled.  Nevertheless, it is a good global representative value at initial times to give an indication of whether or not the cloud will collapse, and on what timescales when comparing models with different mass-to-flux ratios.

%V2: Assuming that the initial cloud has uniform density, the mass-to-flux ratio decreases on concentric spheres within the cloud of smaller and smaller radii, $r$ (i.e. for $r < R$; $\mu(r) \propto r$).  Thus, near the centre of the cloud, the mass-to-flux ratio may reach sub-critical values which would naively suggest that those spheres are supported against gravitational collapse.  However, this assumption neglects the mass outside the small sphere, which will necessarily collapse into the smaller sphere, assuming the cloud is globally super-critical.  Thus, although a useful global representation, the mass-to-flux ratio may provide misleading interpretations if used within the cloud itself, or if used for non-uniform cloud cores.

We will begin the review by describing the observational motivations in Section~\ref{sec:obvs}, followed by a description of ideal MHD in the introduction to Section~\ref{sec:imhd}.  Our focus will then shift to numerical models, where we demonstrate the magnetic braking catastrophe (Section~\ref{sec:imhd:idealized}), followed by attempts to prevent it while still keeping the ideal MHD approximation (Sections~\ref{sec:imhd:mis} and \ref{sec:imhd:turb}).  We will then introduce non-ideal MHD (Section~\ref{sec:nimhd}), and show the recent success of those simulations in preventing the magnetic braking catastrophe.  We will conclude in Section~\ref{sec:conc}.

%------------------------------------------------------------------------------------------------------------------------------------------------------------------------------------------------------------------------------------------------
\section{Observational Motivations}
\label{sec:obvs}

The notion of a magnetized interstellar medium (ISM) dates back more than half a century, to at least the detection of polarized starlight \citep{Hall1949,Hiltner1949} and its interpretation as coming from the absorption of the unpolarized starlight by magnetically aligned grains in the foreground medium (\citealp{DavisGreenstein1951}; see \citealp{AnderssonLazarianVaillancourt2015} for a recent review). With the advent of observational capabilities, the magnetic fields in the ISM in general, and star-forming molecular clouds in particular, are becoming increasingly better characterized. For example, the PLANCK all-sky survey of the dust polarization leaves little doubt that a rather ordered magnetic field component exists in all nearby clouds \citep{Planck2015}, as reviewed by H. B. Li in this volume. Observations have also revealed the prevalence of the magnetic field on the smaller scales of individual cores of molecular clouds and protostellar envelopes, as reviewed by Pattle et al., Crutcher \& Kemball, and Hull \& Zhang, in this volume. 

\begin{figure}
\begin{center}
\includegraphics[width=\columnwidth]{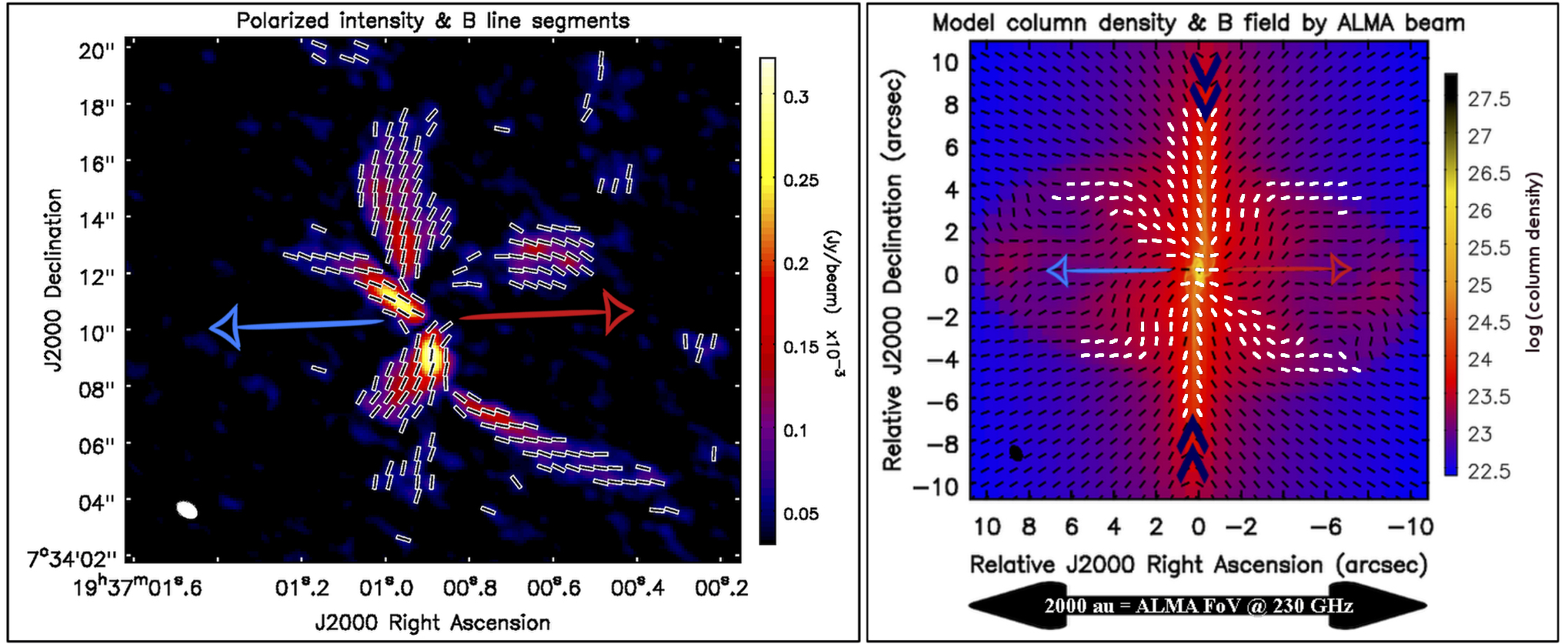}
\caption{An example of the magnetic field traced by dust polarization around an observation (left-hand panel) and a numerical model (right-hand panel) of the solar-type Class 0 protostar B335. The background image in the right-hand panel is the ALMA polarized dust continuum emission, and the superimposed lines infer the magnetic field orientations (i.e. the polarization angle rotated by $90^{\circ}$).    This figure is inspired by figs. 1 and 3 of \citet{Maury+2018}, and was created by A. J. Maury for this publication.}
\label{fig:B335}
\end{center}
\end{figure}

As an illustration, we show in Fig.~\ref{fig:B335} the dust polarization detected with the Atacama Large Millimeter/submillimeter Array (ALMA) around the Class 0 protostar B335 \citep{Maury+2018}. The polarization orientations are rotated by $90^\circ$ to trace the magnetic field directions in the plane of the sky. It is immediately clear that not only a magnetic field is present on large scale, but also it shows coherent structures. In particular, the (projected) field appears to be significantly pinched near the equator of the system, as defined by the bipolar molecular outflows. The pinch is direct evidence that the magnetic field is interacting with the envelope material, through a magnetic tension force. Whether such a magnetic force is strong enough to affect the dynamics of the core collapse and especially disc formation is the question that we seek to address in this article. 

There is some indirect evidence that magnetic fields may play a role in disk (and binary) formation. For example, \citet{Maury+2010} concluded that core collapse models with a relatively strong magnetic field are more consistent with their IRAM-PdBI observations of Class 0 protostellar systems than their hydrodynamic (non-magnetic) counterparts. In the particular case of B335, the specific angular momentum is observed to decrease rapidly towards the central protostar, with a rotationally supported disk (if present) smaller than $\sim 10$~au \citep{Yen+2015}. The decrease in specific angular momentum and small disk could result naturally from the braking by a magnetic field, which has now been mapped in detail with ALMA \citep{Maury+2018}. In addition, there is some tentative evidence that protostellar sources with misaligned magnetic field and rotation axis (inferred from outflow direction) tend to have larger disks \citepeg{Seguracox+2016}, which is consistent with magnetized disk formation simulations \citepeg{HennebelleCiardi2009,JoosHennebelleCiardi2012,KrumholzCrutcherHull2013,LiKrasnopolskyShang2013}.

Finding evidence for the magnetic field on the disc scale is more challenging. Spatially resolved dust polarization has been detected in discs around a number of young stellar objects, using the Submillimeter Array \citep[SMA; e.g.][]{Rao+2014}, the Combined Array for Research in Millimeter-wave Astronomy \citep[CARMA; e.g.][]{Stephens+2014,Seguracox+2015}, the Very Large Array \citep[VLA; e.g.][]{Cox+2015,Liu+2016}, and especially ALMA \citep[e.g.][]{Kataoka+2017,Stephens+2017,Lee+2018,Girart+2018,Hull+2018,Sadavoy+2018,Harris+2018,Cox+2018,Alves+2018,Bacciotti+2018,Dent+2019}. However, with the exception of BHB 07-11 \citep{Alves+2018} and possibly a few other cases, the majority of the sources do not show any evidence for dust grains aligned by the generally expected toroidal magnetic fields; their polarization patterns are better explained by dust scattering instead \citep{Kataoka+2015,Kataoka+2016polarisation,Yang+2016hltau,Yang+2016ngc,Yang+2017}.  The reader is referred to Hull \& Zhang's article in this volume for a more detailed discussion. In any case, whether and how the disc is connected to the protostellar envelope through a magnetic field remain to be determined observationally. 

%------------------------------------------------------------------------------------------------------------------------------------------------------------------------------------------------------------------------------------------------
\section{Disc Formation in the Ideal MHD Limit}
\label{sec:imhd}

\begin{figure}
\begin{center}
\includegraphics[width=.80\columnwidth]{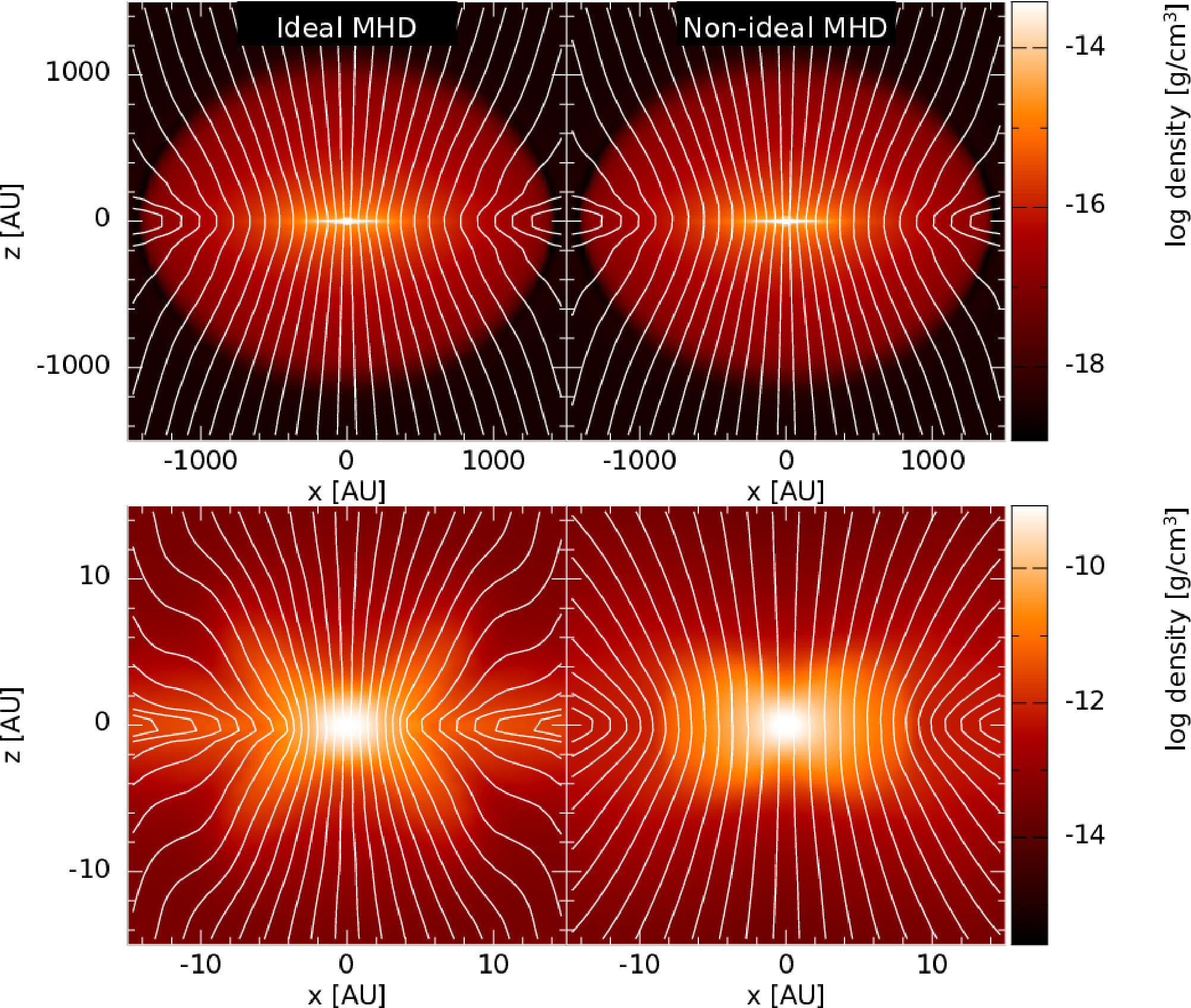}
\caption{The magnetic field lines superimposed on a density slice during the first hydrostatic core phase for ideal and non-ideal MHD simulations.  The initial mass-to-flux ratio is five times the critical value (i.e. $\mu_0=5$).  In ideal MHD, the magnetic field lines are dragged inwards as the cloud core collapses, creating the characteristic hour-glass shape.  On the large scale, the non-ideal effects have minimal effect on the strength and structure of the magnetic field, whereas on the small scale, the neutral particles flow through the magnetic field lines to form the first hydrostatic core, while preventing the magnetic field lines from becoming pinched and preventing a strong magnetic field from building up in the core (bottom).  These images are inspired by fig. 2 of \citet{PriceBate2007} and fig. 1 of \citet{BateTriccoPrice2014}.}
\label{fig:Bfield}
\end{center}
\end{figure}

The simplest approximation when modelling magnetic fields is to use ideal MHD, where it is assumed that the gas is sufficiently ionized such that the magnetic field is well coupled to the bulk neutral gas. In this approximation, the induction equation is given by
\begin{flalign}
\label{eq:imhd}
\frac{\partial\bm{B}}{\partial t} &=  \bm{\nabla} \times \left(\bm{v}\times \bm{B}\right), &
\end{flalign}
where $\bm{v}$ is the gas velocity and $\bm{B}$ is the magnetic field.  Since the gas is tied to the magnetic field lines, the lines are dragged in as the gas collapses (assuming $\mu_0 > 1$), causing a characteristic hour-glass shape; see the left-hand column of Fig.~\ref{fig:Bfield} for numerical results, which nicely complement the observational results in Fig~\ref{fig:B335}. 
This pinching effect becomes less prominent as the magnetic field becomes stronger, since the stronger field is harder to bend \citep[see fig.~2 of][]{PriceBate2007}.  If the magnetic field were to be dragged all the way into the central stellar object, then the stellar field strength would be    millions of gauss, which is much higher than the kilo-gauss field typically observed in young stars. This is a manifestation of the so-called ``magnetic flux problem'' in star formation \citep{BabcockCowling1953,MestelSpitzer1956,Shu+2006}\footnote{Given the focus of this review, the magnetic flux problem will not be addressed here; see \citet{WursterBatePrice2018ff} for a recent  discussion.}.  

In purely hydrodynamics simulations, large protostellar discs can form due to conservation of angular momentum.  In the presence of magnetic fields, angular momentum can be efficiently transported away from the collapsing central region \citep[e.g.][]{MestelSpitzer1956,MouschoviasPaleologou1979,MouschoviasPaleologou1980,BasuMouschovias1994,BasuMouschovias1995b,MellonLi2008}, and not enough angular momentum remains for a rotationally supported disc to form.  This is the magnetic braking catastrophe, as first demonstrated by \citet[][see also the pioneering work by \citealp{Tomisaka2000}]{AllenLiShu2003}:  Rotationally supported discs do not form in idealized numerical simulations in the presence of magnetic fields of realistic strengths.  Analytical studies by  \citet{JoosHennebelleCiardi2012} estimated that $\mu \le 10$ should be enough to suppress disc formation.  

There have been many numerical simulations of disc formation under the assumption of ideal MHD.  Most simulations are initialized with a rotating spherical cloud core which is threaded with a magnetic field that is parallel to the rotation axis  (Section~\ref{sec:imhd:idealized}).  However, molecular clouds contain turbulent flows \citep[e.g.][]{HeyerBrunt2004}, which form large scale structures \citep[e.g.][]{PadoanNordlund2002,MckeeOstriker2007,Wardthompson+2010}, and it is these chaotic structures that can collapse to form cores that ultimately collapse to form stars and protostellar discs.  Thus, a more realistic scenario is that the magnetic fields are initially misaligned with the rotation axis (see Section~\ref{sec:imhd:mis}), or the velocity field initially contains a significant turbulent component (see Section~\ref{sec:imhd:turb}).

%------------------
\subsection{Idealized Initial conditions}
\label{sec:imhd:idealized}

The simplest and most common initial condition for disc formation from a collapsing molecular cloud core is to thread a magnetic field parallel to the rotation axis of a spherical core that is in solid-body rotation; the initial magnetic field strength is characterized by the initial mass-to-flux ratio, $\mu_0$, for the core as a whole.
The early ideal MHD simulations were performed under the assumption of an isothermal or barotropic equation of state and were performed in two-dimensional \citepeg{AllenLiShu2003,MellonLi2008} or three-dimensional \citepeg{MachidaTomisakaMatsumoto2004,PriceBate2007,HennebelleFromang2008,DuffinPudritz2009,MachidaInutsukaMatsumoto2011,Zhao+2011,Seifried+2012,Santoslima+2012}.
 Later studies were radiative three-dimensional calculations that included simplified ideal magnetic fields calculations \citepeg{Boss1997,Boss1999}, or solved the complete MHD equations \citepeg{Boss2002,Boss2005,Boss2007,Boss2009,Commercon+2010,Tomida+2010rmhd,Tomida+2013,BateTriccoPrice2014}.  Subsequent studies included radiation and ideal magnetic fields as part of a parameter study 
\citep[e.g.][]{Tomida2014,TomidaOkuzumiMachida2015,Tsukamoto+2015oa,WursterBatePrice2018sd,Vaytet+2018,WursterBatePrice2018hd,WursterBatePrice2018ff}.  When using moderate to strong magnetic fields, these studies all found efficient magnetic braking, and none of them formed a protostellar disc.  

\begin{figure}
\begin{center}
\includegraphics[width=\columnwidth]{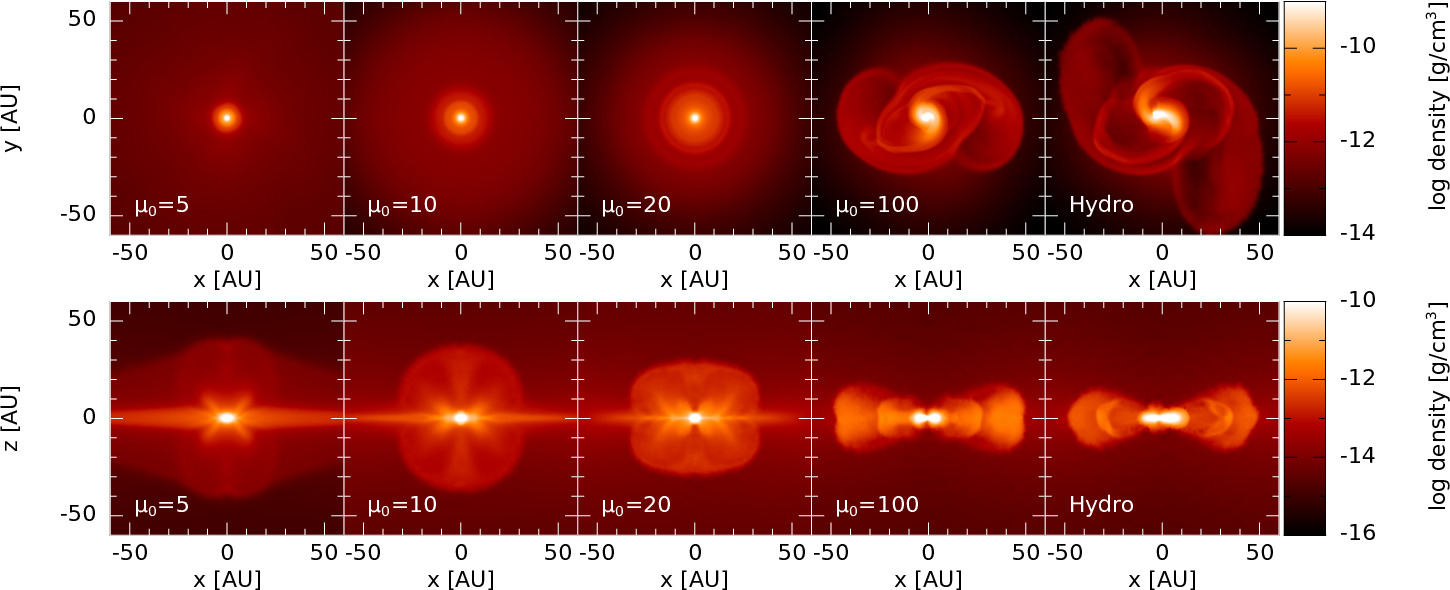}
\caption{The face-on (top row) and edge-on (bottom row) gas density in a slice through the centre of the first hydrostatic core for ideal MHD models of decreasing magnetic field strength (increasing $\mu_0$; left to right) and a pure hydrodynamics model.  Gravitationally unstable discs form for $\mu_0 > 20$, whereas only pseudo-discs only form in the remaining models.  This is inspired by fig. 4 of \citet{BateTriccoPrice2014}, and was created for this publication using the data from \citet{BateTriccoPrice2014}.}
\label{fig:idealcollapse}
\end{center}
\end{figure}

For a demonstration of the magnetic braking catastrophe, \citet{BateTriccoPrice2014} simulated four magnetized models and one hydrodynamical model.  Fig.~\ref{fig:idealcollapse} shows the face-on and edge-on gas densities in a slice through the first hydrostatic core, and these figures are representative of ideal MHD models in the literature.
With weak or no magnetic fields ($\mu_0 = 100$, Hydro), the first core is rotating quickly enough and is massive enough to become bar unstable and forms a gravitationally unstable disc that is dominated by spiral arms \citep[e.g.][]{Bate1998,SaigoTomisaka2006,SaigoTomisakaMatsumoto2008,MachidaInutsukaMatsumoto2010,Bate2010,Bate2011}.  

As analytically predicted by \citet{JoosHennebelleCiardi2012}, there are no discs in the models with $\mu_0 \le 10$, however, pseudo-discs do form; a pseudo-disc is an over-density of gas around a protostar that is not centrifugally supported, not in equilibrium, and is resulted from the anisotropy of the magnetic support against gravity \citep{GalliShu1993,LiShu1996}, although, throughout the literature, authors use this term to refer to a variety of disc-like structures.  The pseudo-discs in \citet{BateTriccoPrice2014} do not increase in size, nor do they ever become Keplerian discs.  This study clearly demonstrates the magnetic braking catastrophe, at least up to the formation of the first hydrostatic core.  

When considering the long term evolution of the system, discs may yet form.  In their ideal MHD simulations, \citet{MachidaHosokawa2013} find that discs form in their models by the end of the Class 0 phase, and increase in mass into the Class I phase.  As the envelope is depleted, the magnetic braking becomes less efficient, which allows these discs to form as speculated earlier by \citet[][see also Section~\ref{sec:nimhd:oa}]{MellonLi2008}; when their strongly magnetized models end in the Class I phase, the discs have masses \sm40 per cent of the mass of the protostar itself.

This leads to the open question of when protostellar discs form.  If they form in later stages (e.g. Class I or II), then there may be no magnetic braking catastrophe in the numerical simulations; if they form early in the Class 0 phase, then the catastrophe persists, and one must go beyond the idealized initial conditions to form a discs if the magnetic field is strong and well-coupled to the gas.  Future observations are required to determine when in the star formation process its protostellar disc forms.

%------------------
\subsection{Misaligned magnetic fields}
\label{sec:imhd:mis}

There have been several studies investigating the impact of misaligned magnetic fields on the formation of discs \citepeg{MatsumotoTomisaka2004,MatsumotoNakazatoTomisaka2006,Machida+2006,HennebelleCiardi2009,JoosHennebelleCiardi2012,KrumholzCrutcherHull2013,LiKrasnopolskyShang2013,LewisBatePrice2015,LewisBate2017}.  Similar to the literature, we define the angle $\theta$ such that the angular momentum $\bm{J}$ and magnetic field $\bm{B}$ vectors are parallel and aligned when $\theta\equiv0^\circ$.  The components of the angular momentum that are parallel and perpendicular to the magnetic field are $\bm{J}_\parallel = \left|\bm{J}\cdot\bm{B}\right|/ \left|\bm{B}\right|$ and $\bm{J}_\perp = \left|\bm{J}\times\bm{B}\right|/ \left|\bm{B}\right|$, respectively.

Two-dimensional analytical models of collapsing cylinders by \citet{MouschoviasPaleologou1979} found that magnetic braking can reduce the angular momentum of a cloud by a few orders of magnitude if \thetanz.  All other parameters being the same, this indicates that systems with \thetaz \ are more likely to form discs than their \thetanz \ counterparts.  However, this pioneering work did not include the gravitational collapse, which can modify the magnetic field configuration and affect the braking efficiency.

The results of \citet{MouschoviasPaleologou1979} were later confirmed by the three-dimensional models of \citet{MatsumotoTomisaka2004}.  In these models, the perpendicular component of the angular momentum, $\bm{J}_\perp$, decreased faster than the parallel component, indicating that magnetic braking was more efficient for the perpendicular component.  This component decreased rapidly and by a few orders of magnitude in their models with $\theta = 45$ and $90^\circ$; the component $\bm{J}_\parallel$ decreased only by a factor of a few in their models with $\theta = 0$ and $45^\circ$ (see their fig.~4).  These results broadly agree with the parameter study by \citet{Machida+2006}, who also find that magnetic braking acts primarily on the component perpendicular to the rotation axis. They conclude that discs form more easily when \thetaz.

Several studies, however, reach the opposite conclusion: Discs form more easily when \thetanz.   \citet{JoosHennebelleCiardi2012} find that massive discs form in all of their misaligned models, requiring as little as $\theta=20^\circ$ to allow a massive disc to form.  The exceptions are their models with the strongest magnetic field strength, $\mu_0=2$, in which discs never form, independent of $\theta$.  As the evolution progresses, the pseudo-discs continue to accrete, increasing both their mass and angular momentum; more massive discs form for larger $\theta$, and faster rotating discs form for weaker magnetic fields (larger $\mu_0$).

Except in the case of very strong magnetic fields, \citet{LiKrasnopolskyShang2013} find the initially misaligned magnetic field allows rotationally supported discs to form in the dense cores, even when no discs form in the aligned models.  In their models, the magnetic field lines are wrapped into a snail-shaped curtain when \thetanz, and this configuration hinders outflows.  With negligible outflows, the angular momentum remains near the protostar, allowing the formation of the disc.

In \citet{LewisBate2017}, the  pseudo-disc increases in size and forms larger arms as the misalignment increases since the gas can easily flow along the horizontal magnetic field component. For $\theta = 20$ and $45^\circ$, the pseudo-discs are warped such that the inner regions are perpendicular to the rotation-axis, while the outer regions are perpendicular to the magnetic field.

In summary, misalignment between the rotation axis and the magnetic field lines may promote or hinder the formation of rotationally supported discs.  \citet{Machida+2006} found that for slow rotators, magnetic braking aligns the rotation axis and the magnetic field (in agreement with \citealt{LewisBate2017} if comparing the outer parts of the discs), and for fast rotators, the magnetic field aligns through a dynamo action.  This suggests that the effect of misalignment may, at least in part, be a result of the initial conditions.  Thus, at the time of writing, the effect of misalignment on disc formation is inconclusive.

%------------------
\subsection{Turbulent initial conditions}
\label{sec:imhd:turb}

There are several studies of disc formation in massive turbulent magnetized molecular clouds \cite[$M > 100$\Msun; e.g.][]{Santoslima+2012,Santoslima+2013,Seifried+2012,Seifried+2013,Myers+2013,Li+2014,Fielding+2015,GrayMcKeeKlein2018}, as well a number of studies that begin from turbulent, low-mass cores \cite[$M < 10$\Msun; e.g.][]{MatsumotoHanawa2011,Joos+2013,Li+2014,MatsumotoMachidaInutsuka2017,LewisBate2018}; for scales consistent with this review, we focus on the latter studies.  These low-mass simulations reach contradicting results, with some studies suggesting increased turbulence promotes disc formation \citep{Joos+2013,Li+2014}, while others suggest it hinders discs formation \citep{MatsumotoHanawa2011,MatsumotoMachidaInutsuka2017,LewisBate2018}.

\begin{figure}
\begin{center}
\includegraphics[width=\columnwidth]{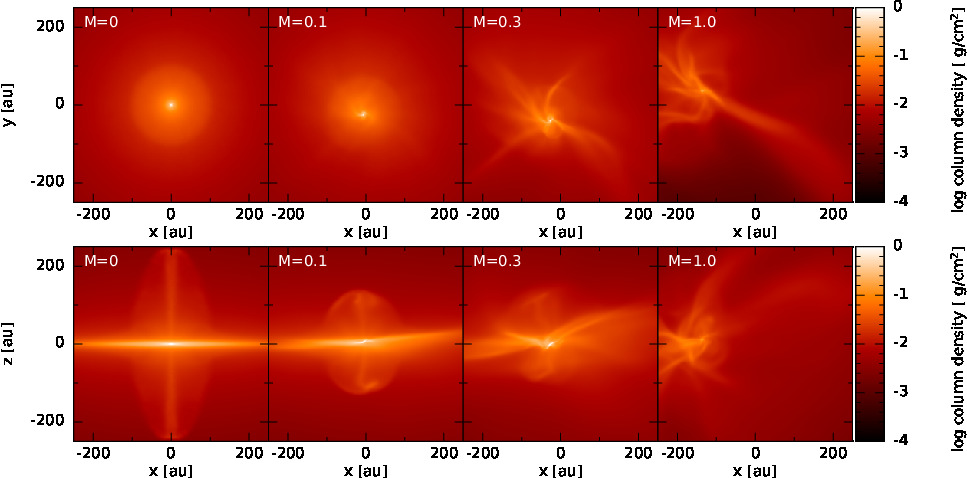}
\caption{The face-on (top row) and edge-on (bottom row) gas column density projections showing disc formation in models in a relatively strong ($\mu_0=5$) magnetic field with a turbulent velocity field imposed onto a solid-body rotation.  The Mach number is shown in each panel.  The figures are taken $\sim$500~yr after sink formation (i.e. when the maximum density has reached $\rho_\text{max} = 10^{-10}$ g~cm$^{-3}$).  Increasing turbulence in these models hinders disc formation.  These are inspired by fig.~7 of \citet{LewisBate2018}, but are from lower resolution models.}
\label{fig:turb}
\end{center}
\end{figure}

Fig.~\ref{fig:turb} illustrates the effect of how increasing the Mach number, $\mathcal{M}$, of the turbulent velocity field imposed on a slowly rotating, pre-stellar cloud core hinders disc formation.
As the Mach number is increased, the resulting pseudo-disc is smaller, and the rotating gas becomes less Keplerian.  For $\mathcal{M}=1$, the system is disrupted and no pseudo-disc forms.  In the most turbulent model, the initial ratio of turbulent to rotational energy is $E_\text{turb}/E_\text{rot} = 26$, and \citet{LewisBate2018} argue that $E_\text{turb}/E_\text{rot} \lesssim 1$ is required for the formation of a pseudo-disc.  By increasing the initial rotation such that $E_\text{turb}/E_\text{rot} = 1.6$, they form a disrupted pseudo-disc, while increasing it such that $E_\text{turb}/E_\text{rot} = 1.06$, they form a slowly rotating pseudo-disc.

On slightly larger spatial scales before the first hydrostatic core forms, \citet{MatsumotoHanawa2011} find that models without turbulence produce axisymmetric oblate or prolate clouds (depending on initial mass).  As the turbulence is increased, the clouds become more chaotic and disrupted.  As the gravitational collapse continues, each model eventually forms a spherical first core surrounded by a disc-like envelope (with the exception of one model with weak magnetic fields and moderate turbulence).

Following the long-term evolution of their turbulent models, \citet{MatsumotoMachidaInutsuka2017} formed a disc in each model, and the disc mass and radius increased with time.  In agreement with non-turbulent studies, they consistently found larger discs in models with weaker magnetic fields.  However, they also consistently found larger discs in models with weaker turbulence (all other parameters being held constant).  Thus, they concluded, turbulence hindered disc formation.  In their strongest magnetic field model, the disc radii and masses were nearly indistinguishable between their two turbulent models ($\mathcal{M} = 0.5,1$; see their figs.~5 and 6), suggesting that at these magnetic field strengths, the strength of turbulence played a secondary role in the cloud's evolution.

\begin{figure}
\begin{center}
\includegraphics[width=0.8\columnwidth]{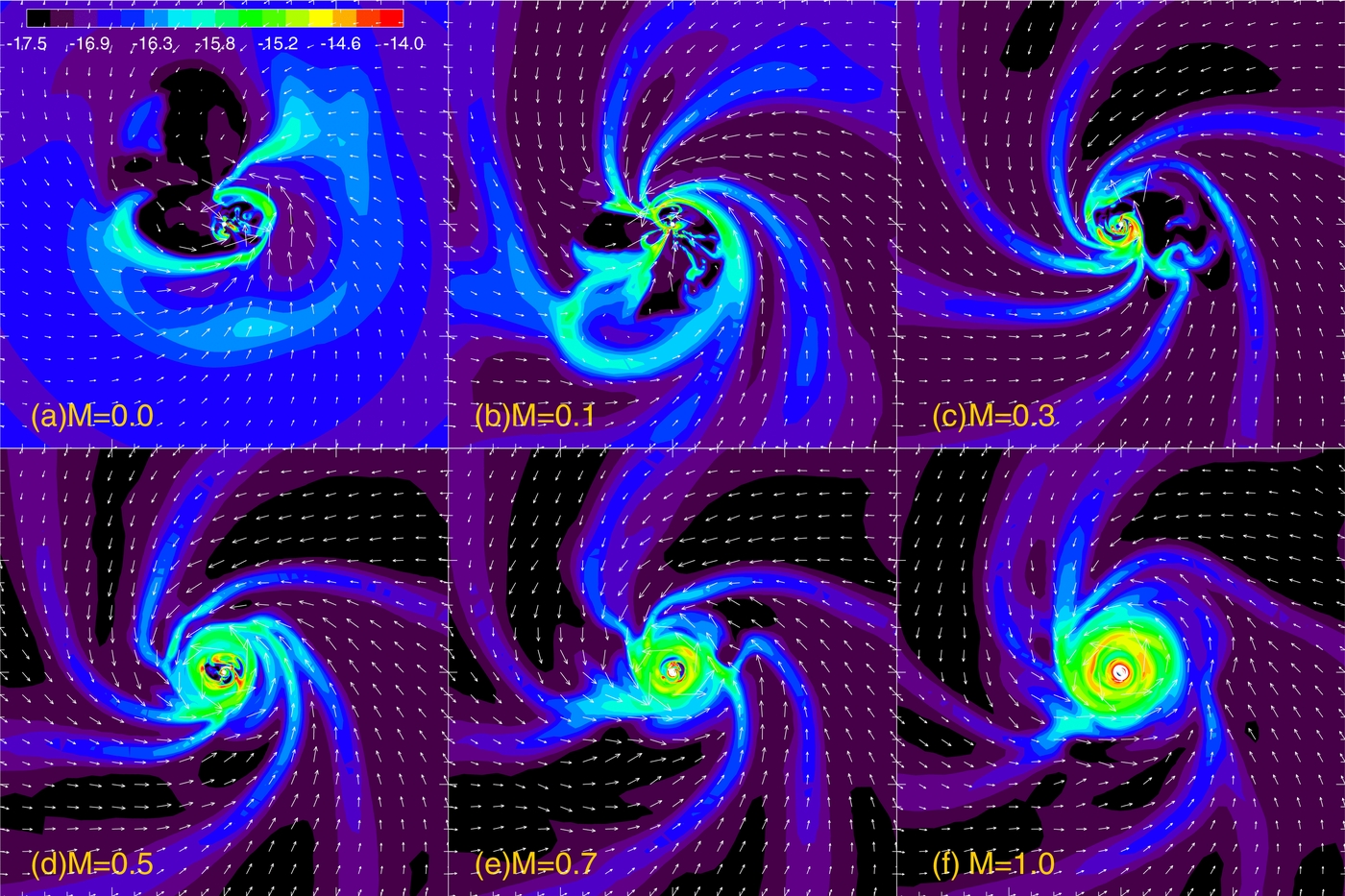}
\caption{The gas density (colour) and velocity (vectors) in the mid-plane for models with increasing Mach number (printed in the bottom left-hand corner of each frame).  Each frame is \sm1400~au on each side.  In this suite of simulations, increasing the Mach number promotes disc formation.  This is fig.~2 of \citet{Li+2014}.  \textsuperscript{\textcopyright}~AAS. Reproduced with permission.}
\label{fig:turb:promote}
\end{center}
\end{figure}

Contrary to the above, Fig.~\ref{fig:turb:promote} illustrates the effect of how increasing the Mach number, $\mathcal{M}$ promotes disc formation.  In the models of \citet{Li+2014}, a disc-like structure begins to form at $\mathcal{M}=0.5$, however, it is still partially disrupted.  At larger Mach numbers, the disc becomes more prominent, and for $\mathcal{M}=1$, has a Keplerian rotational profile.  They conclude that the promotion of disc formation is a result of the warping of the pseudo-disc and the magnetic decoupling-triggered reconnection of the severely pinched field lines near the central object. 

\citet{Joos+2013} presented a suite of models, and found massive discs in all their simulations with weak magnetic fields, and very small discs in their strongly magnetised models; at both strong and weak magnetic field strength, the disc growth rate is approximately independent of the Mach number.  For moderate magnetic field strengths, the disc growth rates are dependent on the Mach number such that a \sm0.6\Msun \ disc  forms in the same length of time it takes to form a \sm0.4\Msun \ disc in a laminar model (the model has an initial mass of 5\Msun).  In these models, the turbulence diffuses the magnetic field out of the central region, generating an effective magnetic diffusivity prompted by magnetic reconnection, and hence weakening the magnetic field \citep[see also][]{Weiss1966,Santoslima+2012}.  Turbulence also induces a misalignment between the rotation axis and the magnetic field \citep[see also][]{Seifried+2012,GrayMcKeeKlein2018} of $20-60^\circ$, which reduces the magnetic braking.  The effect of these two mechanisms is to allow for larger discs to form.

In summary, turbulence can hinder or promote disc formation.  Thus, as with the studies of initial magnetic field alignment, initial conditions will likely play an important role in determining the outcome.

%------------------------------------------------------------------------------------------------------------------------------------------------------------------------------------------------------------------------------------------------
\section{Non-Ideal MHD and Disc Formation} 
\label{sec:nimhd}

It is well known that the dense, star-forming, cores of molecular clouds are lightly ionized \citep{BerginTafalla2007}, with detailed models finding ionization fractions as low as $n_\text{e} /n_{\text{H}_2} = 10^{-14}$ \citep{NakanoUmebayashi1986,UmebayashiNakano1990,NishiNakanoUmebayashi1991,NakanoNishiUmebayashi2002}.  The low ionization level means that the magnetic field is no longer perfectly coupled to the bulk neutral material, rendering the ideal MHD approximation questionable. A proper treatment of the non-ideal MHD effects is required, including a detailed calculation of the abundances of the electrons, ions and charged dust grains.   

There are several methods of numerically modelling multiple species.  They can be modelled explicitly \citep[e.g.][]{InoueInutsukaKoyama2007,InoueInutsuka2008,InoueInutsuka2009}, where each species has its own continuity and momentum equation.  In this method, the species interact directly with each other through terms in the momentum and energy equations.  Electrons are not explicitly treated because their mass is much less than those of other particles.  The induction equation is as given in Eqn.~\ref{eq:imhd}, except that it only includes the velocity of the charged species.

Under the assumptions that mass density is dominated by the neutral mass density, and that collisions occur predominantly between charged species and neutrals, then the inertia and pressure of the charged species and collisions between charged species can be safely neglected \citep[e.g.,][]{OsullivanDownes2006,RodgersleeRayDownes2016}.  In this approximation, each species has its own continuity equation, but only the neutral species has a momentum equation; additional equations are included to govern the interactions (e.g. energy transfer) between the species.  Other studies include dust grains, and include continuity equations for the neutral gas and total grain density but not the charged gas species \citepeg{CiolekMouschovias1993,CiolekMouschovias1994,TassisMouschovias2005a,TassisMouschovias2005b,TassisMouschovias2007a,TassisMouschovias2007b,TassisMouschovias2007c,KunzMouschovias2009,KunzMouschovias2010}.

The continuity equation can be constructed to evolve total mass density rather than species mass density if the strong coupling approximation is invoked; in this approximation, the ion pressure and momentum are negligible compared to that of the neutrals, and the magnetic field and neutral flows evolve on a timescale that is long compared to the timescale of the charged particles \citepeg{WardleKoenigl1993,CiolekMouschovias1994,Maclow+1995,WardleNg1999,ChoiKimWiita2009}.  In approximation, there is one continuity and one momentum equation.  This method is typically used by those studying the formation of protostellar discs \citepeg{KrasnopolskyLiShang2010,TomidaOkuzumiMachida2015,Tsukamoto+2015oa,Tsukamoto+2015hall,WursterPriceBate2016,Masson+2016,Vaytet+2018,WursterBatePrice2018hd}.

Except in the cases where the charged and neutral species interact directly through their own momentum equations, the induction equation is modified to account for the species of different charges, viz.,
\begin{comment}
\begin{flalign}
\label{eq:nimhd}
\frac{\partial\bm{B}}{\partial t} &= \left.\frac{\partial\bm{B}}{\partial t}\right|_\text{ideal} + \left.\frac{\partial \bm{B}}{\partial t}\right|_\text{non-ideal} &\notag \\
&= \bm{\nabla} \times \left(\bm{v}\times \bm{B}\right) - \bm{\nabla} \times \left[\eta_\text{O}\bm{J} + \eta_\text{H} \bm{J}\times\hat{\bm{B}} - \eta_\text{A} \left(\bm{J}\times\hat{\bm{B}}\right)\times\hat{\bm{B}} \right], &
\end{flalign}
where $\bm{J}$ is the current density\footnote{Note that in Section~\ref{sec:imhd:mis}, $\bm{J}$ represented angular momentum.}, and
\end{comment}
\begin{flalign}
\label{eq:nimhd}
\frac{\partial\bm{B}}{\partial t} &= \left.\frac{\partial\bm{B}}{\partial t}\right|_\text{ideal} + \left.\frac{\partial \bm{B}}{\partial t}\right|_\text{non-ideal} &\notag \\
&= \bm{\nabla} \times \left(\bm{v}\times \bm{B}\right) - \bm{\nabla} \times \left\{\eta_\text{O}\left(\bm{\nabla} \times \bm{B}\right) 
                                                                                                                         + \eta_\text{H}\left(\bm{\nabla} \times \bm{B}\right)\times\hat{\bm{B}} 
                                                                                                                         - \eta_\text{A} \left[\left(\bm{\nabla} \times \bm{B}\right)\times\hat{\bm{B}}\right]\times\hat{\bm{B}} \right\}, &
\end{flalign}
where all the micro-physics governing the species properties and interactions is contained within the coefficients, $\eta$.  In the case of, e.g., \citet{OsullivanDownes2006} and \citet{RodgersleeRayDownes2016}, the species densities are taken directly from the continuity equation, whereas sub-grid algorithms are required if not explicitly evolving the density of the charged species or if using the strong coupling approximation.  The three coefficients represent the different regimes of interactions between the neutrals and the charged particles: 
\begin{enumerate}
\item Ohmic resistivity, $\eta_\text{O}$:  ions, electrons and charged grains are completely decoupled from the magnetic field,
\item Hall effect (ion-electron drift), $\eta_\text{H}$: massive particles (ions, charged grains) are decoupled from the magnetic field while electrons remain coupled (i.e. the electrons are frozen into the magnetic field, which drifts through the ions and charged grains), and
\item Ambipolar diffusion (ion-neutral drift), $\eta_\text{A}$: both massive charged particles (ions, charged grains) and electrons are coupled to the magnetic field (i.e. the charged particles are frozen into the magnetic field, which drifts through the neutrals).
\end{enumerate}

Ohmic resistivity and ambipolar diffusion are both diffusive terms, with the associated energy dissipation given by \citep{WursterPriceAyliffe2014}
\begin{flalign}
\label{eq:dudtni}
%\frac{\partial u}{\partial t} &= \left.\frac{\partial u}{\partial t}\right|_\text{ideal} + \frac{J^2}{\rho}\eta_\text{O}+\frac{1}{\rho}\left[J^2 - \left(\bm{J}\cdot\hat{\bm{B}}\right)^2\right]\eta_\text{A}, & 
\frac{\partial u}{\partial t} &= \left.\frac{\partial u}{\partial t}\right|_\text{ideal} + \frac{\left|\bm{\nabla} \times \bm{B\right|}^2}{\rho}\eta_\text{O}+\frac{1}{\rho}\left\{\left|\bm{\nabla} \times \bm{B}\right|^2 
- \left[\left(\bm{\nabla} \times \bm{B}\right)\cdot\hat{\bm{B}}\right]^2\right\}\eta_\text{A}, & 
\end{flalign}
where $u$ is the internal energy.  Physically, these non-ideal processes allow the neutral and selected ionized particles to slip through the magnetic field, which typically leads to a redistribution of the magnetic field lines relative to the bulk neutral matter. In particular, a concentration of matter may not lead to as large an increase in the magnetic field strength as in the ideal MHD limit. 

Ohmic resistivity is typically important at the highest densities, such as the inner midplane regions of the protostellar disc, while ambipolar diffusion typically dominates at relatively low densities, such as the molecular cloud core itself, and the upper and outer regions of the protostellar disc  \citep[e.g.][]{Shu+2006,Wardle2007,MachidaInutsukaMatsumoto2008,WursterBatePrice2018sd}.  Fig.~\ref{fig:Bfield} shows the magnetic field lines on both the molecular cloud core and protostellar disc scales for ideal and non-ideal MHD.  When using ideal MHD, the magnetic field lines are dragged into the centre, causing an enhancement in the magnetic field strength, and creating the expected `hour-glass shaped' field lines.  Since the neutral gas can slip through the magnetic field lines in the non-ideal MHD case, the field lines do not become as pinched, resulting in a weaker central magnetic field strength.

The Hall effect is dispersive rather than dissipative, and typically dominates at the intermediate densities between the two diffusive regimes, including in parts of the dense core and protostellar disc (e.g. \citealp{SanoStone2002a,SanoStone2002b}; see fig.~3 of \citealp{LiKrasnopolskyShang2011} for an illustrative example).  Unlike the other terms, the Hall effect can change the magnetic geometry of the system without any dissipation.  Specifically, it will generate a toroidal magnetic field from a poloidal magnetic field, where the resulting magnetic torques can induce a rotation in an otherwise non-rotating medium (for an example, see fig.~15 of \citealp{LiKrasnopolskyShang2011}; for a sketch of the processes, see fig.~1 of \citealp{Tsukamoto+2017}).  When the polarity of the initial poloidal magnetic field is reversed, the resulting toroidal magnetic field and rotation would also be in the opposite direction.  This has implications if the system is already rotating, since the Hall effect will either transport angular momentum to or from the central region, thus will either increase or decrease the angular velocity of the local gas \citep{Wardle2007,BraidingWardle2012sf}.  Assuming $\eta_\text{H} < 0$ \citep[as is reasonable in protostellar discs;][]{Tsukamoto+2015hall,Marchand+2016,WursterPriceBate2016,Wurster2016}, then the Hall effect will increase the angular momentum in the disc if the magnetic field vector and rotation axis are initially anti-aligned, and will decrease the angular momentum if the two vectors are aligned \citep[e.g.][]{KrasnopolskyLiShang2011,LiKrasnopolskyShang2011,BraidingWardle2012sf,BraidingWardle2012accretion,Tsukamoto+2015hall,WursterPriceBate2016,Tsukamoto+2017,WursterBatePrice2018sd,WursterBatePrice2018hd}.

%------------------
\subsubsection{Calculating the non-ideal MHD coefficients }
\label{sec:nimhd:species}

When using the modified induction equation as given in Eqn.~\ref{eq:nimhd}, the dependencies of the non-ideal coefficients are $\eta \equiv \eta\left(\rho_\text{gas},T_\text{gas},B,n_j,m_j,eZ_j\right)$ for an arbitrary number of species, where $n_j$, $m_j$ and $eZ_j$ are the number density, mass and electric charge of species $j$, and only the charged species (i.e. $eZ_j \ne 0$) contribute to $\eta$; the equations can be found in \citet{WardleNg1999}, \citet{Wardle2007} and other papers, and thus will not be repeated here.  

The species to be included in the calculation must be selected in advance, with the reaction rates between them determined experimentally or estimated theoretically \citep[e.g. as presented in the UMIST Database;][]{Mcelroy+2013}.  The ionization sources must also be selected in advance, with the primary source for disc formation in typical molecular clouds being cosmic rays.  The cosmic ray ionization rate is given by $\zeta_\text{cr} \approx \zeta_\text{cr,0} \exp\left(-\Sigma/\Sigma_\text{cr}\right)$, where $\Sigma_\text{cr}$ is the characteristic column density for the attenuation of cosmic rays and $\zeta_\text{cr,0}$ is the unattenuated cosmic ray ionization rate. The latter has a canonical value of $\zeta_\text{cr,0}=10^{-17}$ s$^{-1}$ for the Milky Way ISM \citep{SpitzerTomasko1968,UmebayashiNakano1981}.  The cosmic ray ionization rate can vary from one region to another in the Galaxy (e.g., it is expected to be higher near a supernova remnant that can accelerate cosmic rays) and be modified by propagation effects, such as magnetic mirroring \citep[e.g.][]{Chandran2000,PadovaniGalliGlassgold2009}.  As a cloud core collapses and a protostellar disc forms, the inner regions of the disc are shielded from cosmic rays \citep{Padovani+2014}, thus other ionization sources become important, such as ionization by X-rays and energetic particles from young stellar objects, and ionization by radionuclide decay.  Canonical X-rays are slightly less energetic and have a shorter attenuation depths than cosmic rays, thus will only affect the surface of the disc \citep[e.g.][]{IgeaGlassgold1999,TurnerSano2008}.  The ionization from radionuclide decay, however, has rates ranging from $\zeta_\text{r} \approx 10^{-23}$ to $10^{-18}$ s$^{-1}$ depending on the ionization source \citepeg{UmebayashiNakano2009,KunzMouschovias2009} and can persist throughout the disc.  Thus, ionization from radionuclide decay may be the dominant ionization source near the midplane of the disc.

Along with ions, molecular clouds include dust grains, which are important in coupling the magnetic field to the gas \citep{NishiNakanoUmebayashi1991}.  These are typically included in the numerical calculations of $\eta$ assuming a single grain population with a fixed radius, fixed bulk density, and a fixed gas-to-dust ratio.  However, the grain size greatly affects the strength of the non-ideal effects; for example, \citet{WursterBatePrice2018ion} showed that smaller grains tend to yield larger non-ideal MHD coefficients (see their fig.~2).  However, molecular clouds do not contain a single grain size; one commonly used size distribution is the MRN \citep*{MathisRumplNordsieck1977} grain distribution, ${\text{d}n_\text{g}(a)}/{\text{d}a} \propto n_\text{H}a^{-3.5},$ where $n_\text{H}$ is the number density of the hydrogen nucleus and $n_\text{g}(a)$ is the number density of grains with a radius smaller than $a$ \citep{DraineLee1984}.  The non-ideal MHD coefficients are sensitive to the upper and lower limits of the distribution. In particular, removing the very small grains of \sm10 to a few 100 \AA \ from the distribution would increase the ambipolar diffusivity by \sm$1-2$ orders of magnitude at number densities below  $10^{10}$~cm$^{-3}$ \citep{Zhao+2016}. 

As the non-ideal coefficients increase in value (either by increasing local density, magnetic field strength, or decreasing ionization rate), the numerical timestep required for numerical stability decreases \citepeg{Maclow+1995,ChoiKimWiita2009,Bai2014,WursterPriceAyliffe2014,WursterBatePrice2018ion}.  The two-dimensional simulations of \citet{KunzMouschovias2009} and \citet{DappBasuKunz2012} include all the above discussed ionization mechanisms to allow for a very low ionisation fraction, and end their calculations prior to the timestep becoming prohibitively small.  However, this is not currently computationally possible for global three-dimensional simulations of disc formation and evolution.  Thus, as a crude approximation, studies typically use $\zeta = \zeta_\text{cr,0}=10^{-17}$ s$^{-1}$.

There are several private algorithms \citep[e.g.][]{NakanoNishiUmebayashi2002,Okuzumi2009,KunzMouschovias2009,DappBasuKunz2012,Tsukamoto+2015oa,Zhao+2016,HiguchiMachidaSusa2018,ZhaoCaselliLi2018} and publicly available codes \citep[e.g][]{Marchand+2016,Wurster2016} that solve chemical networks of varying complexity to calculate the number density of each species, which can then be used to self-consistently calculate the non-ideal MHD coefficients.  These algorithms include ionisation and reconnection amongst the included species (including dust), and ionization from cosmic rays.  The results are expectedly dependent on the complexity of the networks and the input parameters, however, there is broad qualitative agreement among them.  However, a direct comparison is difficult due to the parameter dependence, and even  a single algorithm can produce widely varying results with small changes to its input parameters (e.g., the assumed dust grain properties).

%------------------
\subsection{Numerical Models}
\label{sec:nimhd:nm}

Ohmic resistivity was the first non-ideal effect to be included in analytical and numerical studies in attempts to prevent the magnetic braking catastrophe.  This was followed by ambipolar diffusion, and finally the Hall effect.  As discussed below, various studies reached various conclusions.  However, recent three-dimensional radiation magnetohydrodynamical simulations \citep{Tsukamoto+2015hall,WursterPriceBate2016,Tsukamoto+2017,WursterBatePrice2018hd} have suggested that the Hall effect can prevent the magnetic braking catastrophe.

\subsubsection{Ohmic resistivity and Ambipolar diffusion}
\label{sec:nimhd:oa}

A first attempt to solve the magnetic braking catastrophe was by \citet{Shu+2006}.  They derived the equations governing the gravitational collapse of a molecular cloud core in the presence of a magnetic field, and included the effects of Ohmic diffusion.  They noted that in order to reduce the magnetic field strength near the central stellar object to a value consistent with meteoritic evidence, $\eta_\text{O} \approx 2\times 10^{20}$~cm$^2$ s$^{-1}$ is required, which is a few orders of magnitude larger than estimated from kinetic theory \citep{Shu+2006}.  A similar calculation by \citet{KrasnopolskyLiShang2010} was able to reduce the required value using different assumptions, but $\eta_\text{O}$ remained uncomfortably high.  Recent calculations of  $\eta_\text{O}$ using chemical networks of varying complexity have shown that this value can be reached in the centre of the first hydrostatic core during the first core phase \citepeg{Marchand+2016,Wurster2016}.   However, this high value persists over a very small $\rho-T$ phase-space, and values much lower than this are expected in the protostellar disc and background medium.

Contradicting the results of \citet{Shu+2006} and \citet{KrasnopolskyLiShang2010}, the three-dimensional numerical studies of \citet{InutsukaMachidaMatsumoto2010} and \citet{MachidaInutsukaMatsumoto2011} formed protostellar discs;  the latter formed a disc of $r \gtrsim 100$~au, and their simulations used sink particles and the barotropic equation of state and were evolved until most of the envelope (i.e. the gas that was initially in the cloud core) had been accreted.   As the envelope was accreted, the \alfven \ waves became less efficient at transporting angular momentum from the gas near the protostar to the remaining envelope (simply because there was not much gas remaining in the envelope, as pointed out earlier in \citealp{MellonLi2008}).  With less envelope and hence less magnetic braking, a \sm40~au protostellar disc formed by \sm$10^4$~yr which grew to \sm100~au by \sm$10^5$~yr.  
A similar study by \citet{WursterPriceBate2016} found \sm10~au discs formed after \sm$3\times10^4$~yr; this smaller disc was likely a result of a lower Ohmic resistivity and a larger sink particle.  Even smaller Ohmic-enabled discs were found in the semi-analytic calculations of \citet{DappBasu2010} and \citet{DappBasuKunz2012}. 

Ohmic resistivity is expected to become important at the highest densities, especially late in the star formation process.  However, by this stage, unless $\eta_\text{O}$ was much higher than expected early on or the envelope was mostly depleted, the magnetic field could have already extracted enough angular momentum from the less dense regions to prevent discs from forming.  

This possibility provides a strong motivation to study ambipolar diffusion, which was expected to be important in the early stages of star formation when the density of the molecular cloud core was still relatively low.   This had long been included in cloud core formation models as a method to diffuse the magnetic field to initiate the quasi-static formation in an initially magnetically subcritical cloud and the later collapse of the cloud core \citep[as first demonstrated by][]{Mouschovias1976b,Mouschovias1977,Mouschovias1979b}.  On the smaller scales regarding protostar formation, it was hoped that this process could diffuse out enough of the magnetic field to permit a Keplerian disc to form.  Analytical calculations by \citet{Hennebelle+2016} predicted that an \sm18~au disc should form when accounting for ambipolar diffusion for parameters typical of molecular cloud cores (see their equation~13); this is much smaller than the 100 au-scale discs that would typically form in the absence of a magnetic field \citep[see also eqn~14 of][]{Hennebelle+2016}.

Numerical simulations have not provided a consensus regarding the effect of ambipolar diffusion.  \citet{MellonLi2009} re-performed their ideal MHD study from \citet{MellonLi2008} but included ambipolar diffusion.  Using two-dimensional axisymmetric models, they concluded that ambipolar diffusion alone did not weaken the magnetic braking enough to allow the formation of a disc that is resolvable by their simulations.  Their conclusion held even when using a cosmic ray ionization rate ten times lower than canonical (i.e. making the system more neutral).  However, their spherical-polar numerical domain included a 6.7~au hole at the centre, thus discs below this radius could necessarily not form, and the artificial boundaries may have suppressed disc formation at radii even slightly larger than the boundary \citep[for comment on effect of sink sizes, see][]{MachidaInutsukaMatsumoto2014}.  

In the three-dimensional studies of \citet{Masson+2016}, Keplerian discs formed during the first hydrostatic core phase in their models that included ambipolar diffusion.  Using a mass-to-flux ratio of 5 times critical and an initial rotation of $\beta_\text{rot} = 0.02$ (the ratio of rotational energy to gravitational potential energy), a disc of \sm80~au formed; for similar initial conditions, small, disc-like structures formed at late times when using ideal MHD.  
By inclining the initial magnetic field (c.f. Section~\ref{sec:imhd:mis}) by $\theta = 40^\circ$ in the ideal MHD models, there was negligible effect on disc formation, leading them to conclude that ambipolar diffusion was more important than inclination in terms of disc formation.  In their models with ambipolar diffusion, the discs formed more easily and were more massive in the non-aligned case. 

\begin{figure}
\begin{center}
\includegraphics[width=0.5\columnwidth]{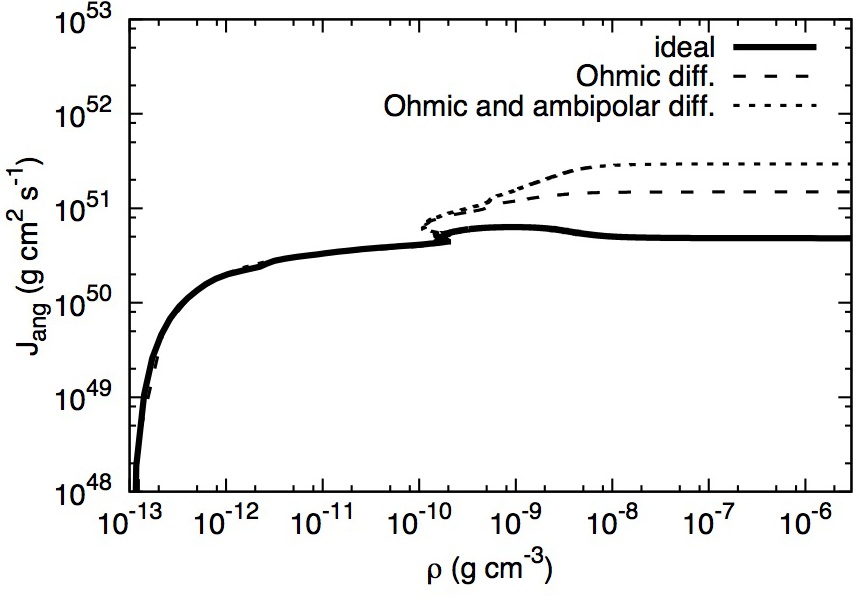}
\caption{The angular momentum in the first hydrostatic core for ideal MHD, Ohmic diffusion, and Ohmic and ambipolar diffusions.  The Ohmic+ambipolar model has enough angular momentum to form a \sm1~au rotationally-supported disc, whereas discs do not form in the other two models.  This figure is inspired by fig. 6 of \citet{Tsukamoto+2015oa}, and was created by Y. Tsukamoto for this publication using the data published in \citet{Tsukamoto+2015oa}.}
\label{fig:oa:L}
\end{center}
\end{figure}

Fig.~\ref{fig:oa:L} shows the angular momentum in the first hydrostatic core (i.e. the gas that would collapse to form a rotationally supported disc if enough angular momentum is present) for three three-dimensional models by \citet{Tsukamoto+2015oa}.  After the first core has formed, there is least angular momentum in the ideal MHD model; there is approximately twice as much angular momentum when Ohmic resistivity is included, and approximately 5 times more when both Ohmic resistivity and ambipolar diffusion are included.  In their model with Ohmic resistivity + ambipolar diffusion, a small \sm1~au disc formed.  At a similar time in the model with Ohmic resistivity + ambipolar diffusion by \citet{TomidaOkuzumiMachida2015}, a centrifugally supported disc of radius \sm5~au formed, although it did not have a Keplerian profile; in their counterpart model that only included Ohmic resistivity, the centrifugally supported disc was \sm1~au.
 
Although the disc in \citet{TomidaOkuzumiMachida2015} is only \sm5~au and remains approximately constant in size throughout their simulation, they expect it to grow with time as the envelope is depleted, as discussed above and in \citet{Tomida+2013}.  By the end of the simulation (approximately 1 year after the formation of the stellar core), the disc is supported by the centrifugal force with substantial contribution from the gas pressure.  At this time, the disc is rotating rapidly enough  that it has triggered the gravito-rotational instability \citep[e.g.][]{Toomre1964,Bate1998,SaigoTomisaka2006,SaigoTomisakaMatsumoto2008} and become non-axisymmetric.

In a followup study, \citet{Tomida+2017} modelled the long term evolution using sink particles.  In agreement with their previous work, the disc stays small at early times due to efficient magnetic braking.  As the disc evolves, magnetic braking becomes less efficient both due to the magnetic field dissipating in the disc and the dissipation of the envelope.  Eventually, the disc becomes unstable, forming an $m=2$ perturbation \citep[see also][]{Hennebelle+2016}.  The gravitational torques become more efficient at transporting angular momentum than the magnetic fields, thus control the future evolution of the disc.  By the end of the Class I phase, the disc radius is in excess of 200~au.

\citet{Vaytet+2018} modelled the collapse to the stellar core including Ohmic resistivity + ambipolar diffusion.  During the first core phase, the gas is funnelled into the core along two dense filaments that arise from an initial  $m=2$ perturbation.  The disc-like structure in their ideal MHD model is `puffier' than their resistive counterpart, but neither model appears to form a disc.  One month after the formation of the stellar core, they find a `second core disc' of radius $r \lesssim 0.1$~au that has a Keplerian velocity profile. 

\citet{Zhao+2016,ZhaoCaselliLi2018} stressed the importance of the grain size on the ambipolar diffusivity \citep[see also][]{DappBasuKunz2012}. They explored a range of grain sizes, initial magnetic field strengths and rotation rates, and found that rotationally supported structures often form early around the stellar seed but disappear at later times. Such structures can persist and grow to sizes of 20-40~au even for a rather strong initial magnetic field corresponding to a dimensionless mass-to-flux ratio of $\mu_0 = 2.4$ when the very small grains are removed from an MRN size distribution. Whether such grains are indeed removed or not, through, e.g., grain coagulation, remain to be determined. 

\subsubsection{Complete non-ideal MHD description}
\label{sec:nimhd:oha}

Including the Hall effect in numerical models becomes computationally expensive due to the small timestep constraint \citep[e.g.][]{SanoStone2002a,Bai2014,Tsukamoto+2015hall,WursterPriceBate2016,WursterBatePrice2018ion}, thus, at the time of writing, there are only a few global two-dimensional \citep{KrasnopolskyLiShang2011,LiKrasnopolskyShang2011} and three-dimensional \citep{Tsukamoto+2015hall,WursterPriceBate2016,Tsukamoto+2017,WursterBatePrice2018sd,WursterBatePrice2018hd,WursterBatePrice2018ff} global disc-formation studies that include this process.

The first simulation to include the Hall effect in a global simulation was \cite{KrasnopolskyLiShang2011}.  This simulation was a two-dimensional axisymmetric Eulerian simulation that included gravity of a pre-formed protostar but no self-gravity amongst the gas.  The Hall coefficient, $\eta_\text{H}$, was set as a constant positive value, which is the opposite sign as found by later studies \citep[e.g.][]{Tsukamoto+2015hall,Marchand+2016,WursterPriceBate2016,Wurster2016}.  They formed rotationally supported discs in all models that included the Hall effect, except for their model with the weakest resistivity.  

\begin{figure}
\begin{center}
\includegraphics[width=\columnwidth]{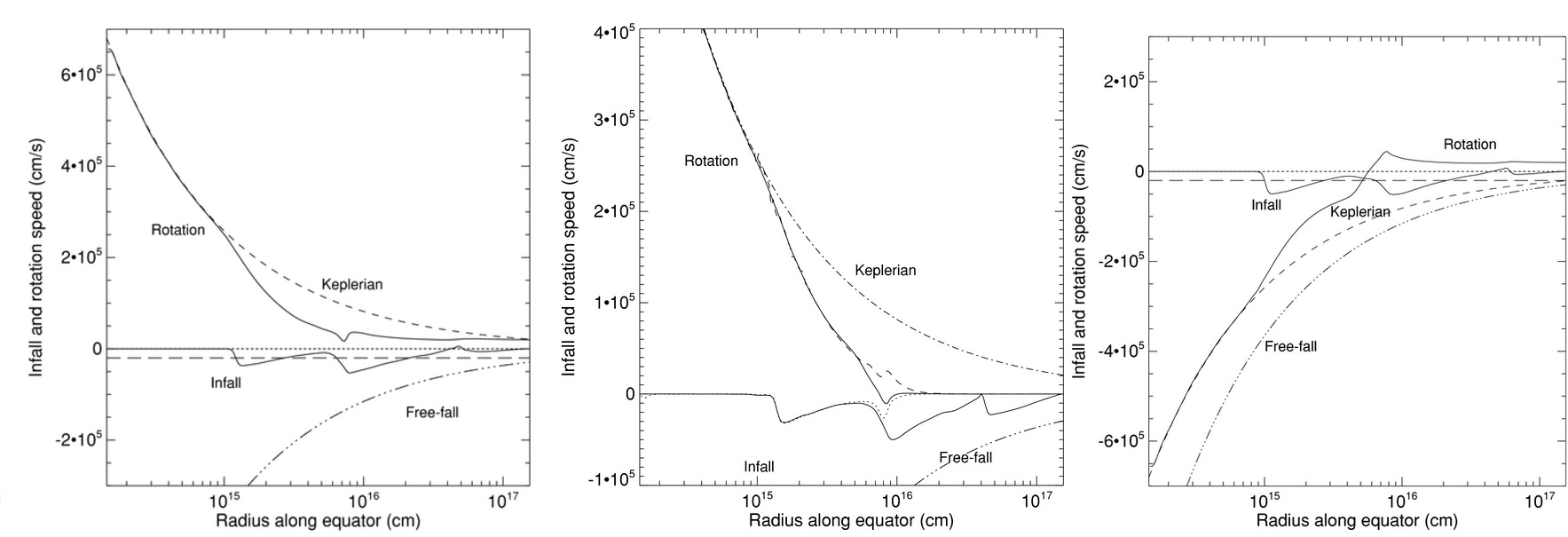}
\caption{Rotational profiles for aligned, no rotation and anti-aligned two-dimensional models.  The initial rotation in the rotating models is $v_\phi = -2\times10^4$~\cms.  The gas in the outer regions is governed by the initial conditions, while Keplerian discs of $r \sim 10^{15}$~cm form in all three models.  These are figs. 1b, 3b and 2b of \citet{KrasnopolskyLiShang2011}. \textsuperscript{\textcopyright}~AAS. Reproduced with permission.}
\label{fig:oha:vphi}
\end{center}
\end{figure}

Fig.~\ref{fig:oha:vphi} shows the rotational profiles for three models in the study by \citet{KrasnopolskyLiShang2011}, where they varied the initial alignment of the rotation and magnetic field vectors.  In the outer regions of their models, the Hall effect is too weak to have any significant effect on the evolution of the gas, thus the gas simply follows the initial velocity profile combined with the effect of gravity.  In the inner region of all three models, the Hall effect spins up the gas to reach a Keplerian profile.  In the aligned model (left panel), the Hall effect contributes to the initial rotation, whereas in the anti-aligned case (right panel), it detracts from it; in the anti-aligned case, the Hall velocity is strong enough to cancel out the initial rotational profile and form a counter-rotating disc with a Keplerian profile.

In their followup study, \citet{LiKrasnopolskyShang2011} self-consistently calculated the value for $\eta_\text{H}$, which now necessarily varied in space and time; $\eta_\text{H} < 0$, and the maximum of its absolute value was lower than the value used in their previous study.  Given their parameter space, no model formed a rotationally supported disc.  The Hall effect did spin up the material of the gas near the protostar, but it was not significant enough to reach Keplerian velocities.

\begin{figure}
\includegraphics[width=1.0\columnwidth]{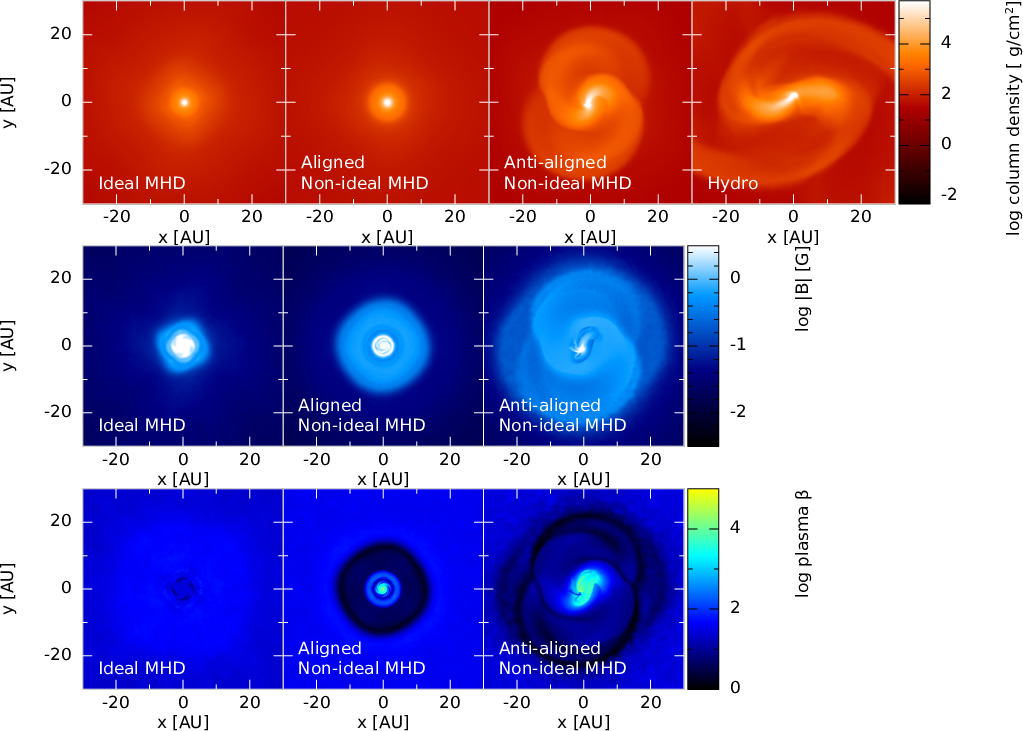}
\caption{The gas column density (top), magnetic field strength (middle) and plasma $\beta$ (bottom) in a slice through the mid-plane of an ideal MHD model, two non-ideal MHD models (Ohmic+ambipolar+Hall; \zetaeq{-17}) with different initial orientations of the magnetic field, and purely hydrodynamics model (top row only).  The snapshots are near the end of the first core phase.  When the rotation and magnetic field vector are aligned, the non-ideal MHD model is similar to the ideal MHD model with no disc, but when the vectors are anti-aligned, a gravitationally unstable disc forms, similar to the hydrodynamics model.  The maximum magnetic field strength is in the disc for the anti-aligned non-ideal model, while it is in the centre of the first core for the remaining two models.  The disc in the anti-aligned model is supported by gas pressure. The figure is adapted/inspired by figs. 1 and 8 of \citet{WursterBatePrice2018hd} and fig.~1 of \citet{Tsukamoto+2015hall}, and was created for this publication using the data from \citet{WursterBatePrice2018hd}.}
\label{fig:oha:rBb}
\end{figure}

\begin{figure}
\begin{center}
\includegraphics[width=0.6\columnwidth]{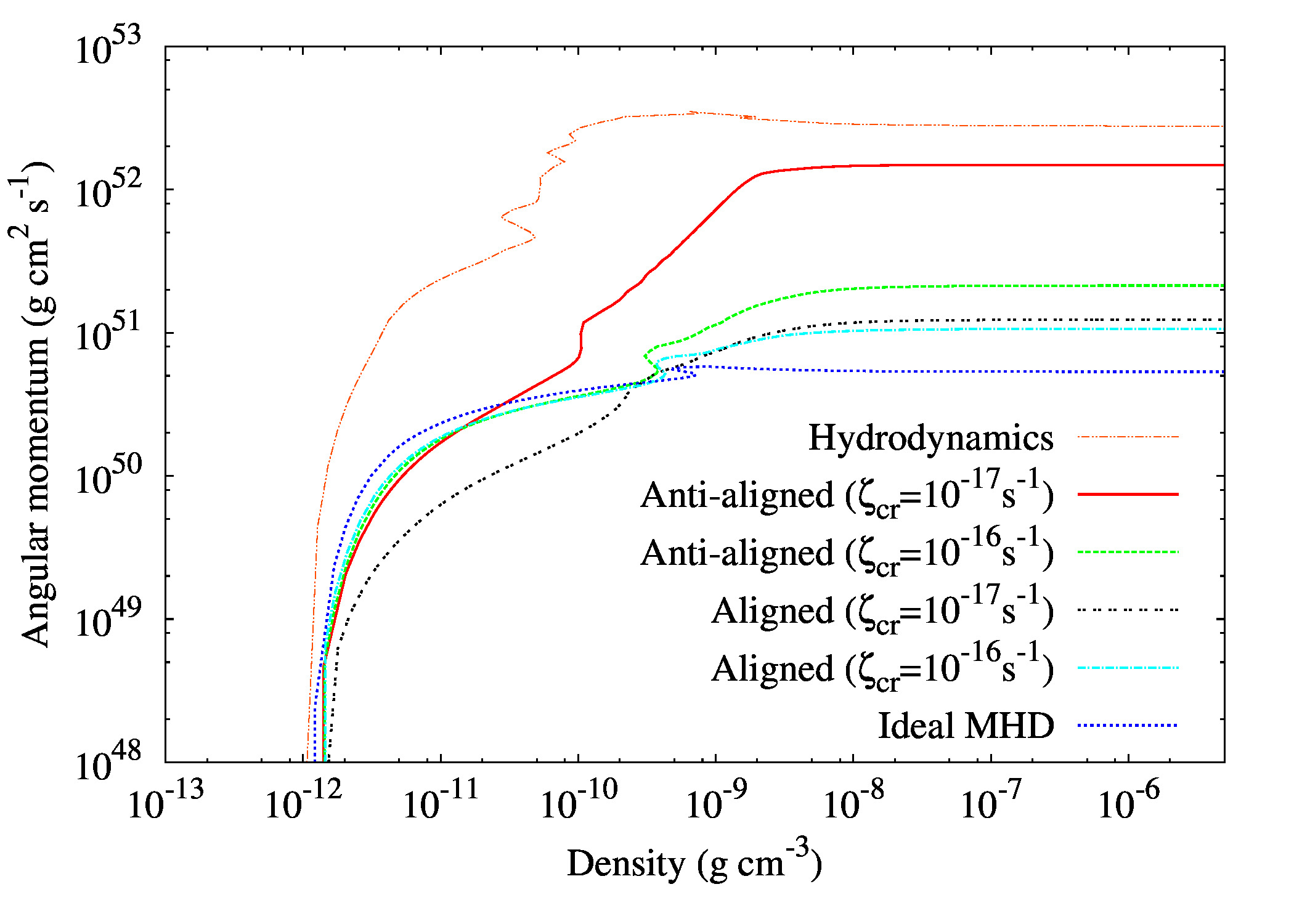}
\caption{The angular momentum in the first hydrostatic core which collapses to form a disc for models with various cosmic ray ionization rates, \zetacr.  Only the purely hydrodynamics and the anti-aligned (\zetaeq{-17}) models form discs, both of which become bar-unstable.  The angular momentum of the ideal MHD model differs slightly from that shown in Fig.~\ref{fig:oa:L} due to different initial conditions.  This is adapted from fig. 2 of \citet{WursterBatePrice2018hd}, and was created for this publication using the data from \citet{WursterBatePrice2018hd}.}
\label{fig:oha:L}
\end{center}
\end{figure}

Subsequent studies led by Tsukamoto and Wurster modelled the three-dimensional collapse of a molecular cloud core through to at least the end of the first core phase.  The models typically included self-consistent calculations of the non-ideal MHD coefficients, included flux limited diffusion and excluded sink particles\footnote{\citet{WursterPriceBate2016} used 6.7~au sink particles together with a barotropic equation of state in their disc formation study using \phantomsph \ \citep{Phantom2018}.}.
When using the canonical, unattenuated cosmic ray ionization rate of \zetaeq{-17}, \citet{Tsukamoto+2015hall} and \citet{WursterBatePrice2018hd} find that rotationally supported discs form if the magnetic field and rotation vectors are anti-aligned, whereas no disc forms if they are aligned.  The top row in Fig.~\ref{fig:oha:rBb} shows the gas column density for ideal MHD, non-ideal MHD and hydrodynamic models in the midplane during the first hydrostatic core phase.  In both studies \citep{Tsukamoto+2015hall,WursterBatePrice2018hd}, the disc has a radius of $r \sim 25$~au, and becomes bar-unstable during the first core phase.  
For comparison, a purely hydrodynamics model formed a disc of \sm60~au, and this disc is formed even earlier during the first core phase due to the lack of magnetic support.  Expectedly, the purely hydrodynamic disc has the largest angular momentum in the first core and/or disc (see Fig.~\ref{fig:oha:L}).  However, the Hall effect in the anti-aligned non-ideal model decreases magnetic braking, permitting the angular momentum to remain in the gas near/in the first core, and hence permitting the disc to form.  The angular momentum in the disc is approximately half that of the purely hydrodynamics model \citep{WursterBatePrice2018hd}.  By aligning the magnetic field and the rotation axis, the angular momentum again decreases by a factor of \sm12, to a value too low for a disc to form.  

The models presented in Figs.~\ref{fig:oa:L} and \ref{fig:oha:L} use slightly different initial conditions, however, both ideal MHD models have similar final angular momenta in their first cores, thus can be reasonably compared.  By including Ohmic resistivity and ambipolar diffusion, the angular momentum in the first core increases to $L_\text{fc} \sim 2.5\times 10^{51}$g~cm$^2$~s$^{-1}$ (dotted line in Fig.~\ref{fig:oa:L}), which is higher than when the three non-ideal terms are added in the aligned orientation (double-dot line in Fig.~\ref{fig:oha:L}).  Therefore, some of the angular momentum gain caused by Ohmic resistivity and ambipolar diffusion is lost by including the Hall effect in an aligned orientation.  As previously discussed, additional angular momentum is gained by the Hall effect in the anti-aligned case.  Summarily, the order of angular momenta in the first core is $L_\text{fc}(\text{ideal MHD}) < L_\text{fc}(\text{Ohmic+ambipolar+Hall; aligned}) < L_\text{fc}(\text{Ohmic+ambipolar}) < L_\text{fc}(\text{Ohmic+ambipolar+Hall; anti-aligned}) < L_\text{fc}(\text{Hydrodynamics})$.

\begin{figure}
\begin{center}
\includegraphics[width=0.6\columnwidth]{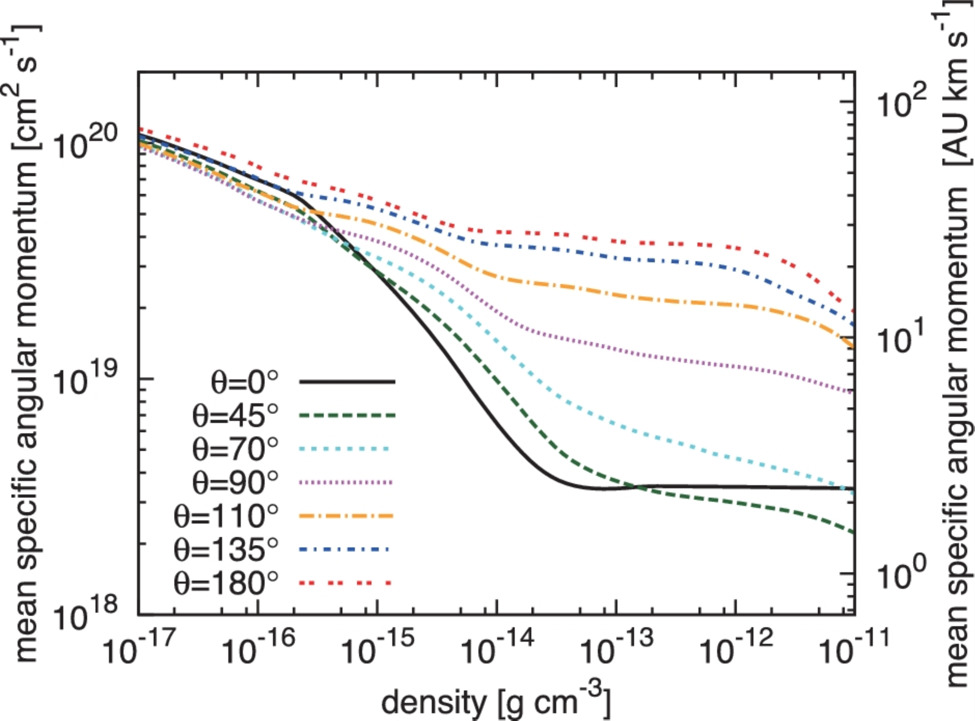}
\caption{The angular momentum in the first hydrostatic core which collapses to form a disc for non-ideal MHD models at various angles $\theta$ between the rotation axis and magnetic field vector; $\theta = 0^\circ$ represents the aligned case.  Except for the $\theta = 90^\circ$ case, the angular momenta are grouped around  $\theta = 0^\circ$ and  $\theta = 180^\circ$, suggesting a bi-modality in disc sizes.  This figure is fig. 6 of \citet{Tsukamoto+2017}, and is reproduced with permission.}
\label{fig:oha:theta:L}
\end{center}
\end{figure}

Given that the anti-aligned non-ideal model formed a large disc while the aligned model did not, \citet{Tsukamoto+2015hall} proposed that there should be a bi-modality in the population of discs around stars - that is, approximately half of the Class 0 objects should have protostellar discs.  However, this assumption was made assuming the magnetic field was either aligned or anti-aligned with the rotation axis.  Since there is observational evidence that the magnetic field appears to be randomly orientated with respect to the rotation axis, at least on the 1000 au scale \citep{Hull+2013}, \citet{Tsukamoto+2017} modelled various angles between the rotation axis and the magnetic field.  The angular momentum in the first core is shown in Fig.~\ref{fig:oha:theta:L} for various initial orientations, and it differs by an order of magnitude between the two extreme angles:  $\theta=0$ and $180^\circ$. Changing the angle by $45^\circ$ from either extreme value has a minimal effect on the angular momentum evolution, and even a change of $70^\circ$ only changes the final angular momentum by a factor of a few.  Thus, the models with $\theta \ne 90^\circ$ have angular momenta evolutions that are similar to either  $\theta = 0$ or  $180^\circ$, suggesting that, even if the initial magnetic fields are randomly orientated with the rotation axis, the bi-modality in disc sizes should exist, with large discs forming for $\theta \gtrsim 100^\circ$ and negligible discs forming for $\theta \lesssim 80^\circ$.  

The magnetic field strength and plasma $\beta$ for an ideal model and two non-ideal MHD models are shown in bottom two rows of Fig.~\ref{fig:oha:rBb}.  As expected, the magnetic field is dragged into the core of the ideal MHD model, thus its first hydrostatic core has a very strong field strength.  However, in the non-ideal cases, the neutrals can slip through the magnetic field, thus a dense core forms, but its central magnetic field strength is weaker than in the ideal MHD model.  Unlike the neutrals, the velocities of the charged particles decrease and they approach the core due to the non-ideal effects.  This allows the charged particles, and the magnetic flux they drag in, to pileup in a torus around the core.  This torus of charged particles expands outwards in radius as more charged particles pile up, further enhancing the magnetic field.  This is a demonstration of a `magnetic wall', which is thoroughly discussed in the two-dimensional models of \citet{TassisMouschovias2005b} and first predicted to exist analytically by \citet{LiMckee1996}.  In the non-ideal MHD models shown above, the magnetic wall occurs at $r\sim1$-3~au from the centre of the core.

In the domain presented in Fig.~\ref{fig:oha:rBb}, plasma $\beta > 1$, indicating that gas pressure is greater than magnetic pressure everywhere.  The disc in the anti-aligned model has plasma $\beta > 100$ due to the rotationally supported disc; in the disc, magnetic diffusion is efficient and the magnetic flux is removed from the region \citep{Tsukamoto+2015hall}.  In the centre of the aligned model, plasma $\beta$ becomes high, but this region is small enough that disc formation is not possible.  Finally, the strong magnetic field in the ideal simulations yields plasma $\beta < 100$ over the entire computational domain, again iterating the importance of magnetic fields in that model.  Similar results with slightly lower values of plasma $\beta$ are reached in \citet[][see their fig. 1]{Tsukamoto+2015hall}, since their initial magnetic field strength is stronger than in the figures presented here.

These two studies discussed here, \citet{Tsukamoto+2015hall} and \citet{WursterBatePrice2018hd}, both modelled the gravitational collapse of a 1M$_\odot$ molecular cloud core, however many of the remaining initial parameters were different, including the initial rotation rate, radius, magnetic field strength, and microphysical properties that controlled the coefficients of the non-ideal MHD terms.  Thus, gravitationally unstable discs forming around the time of the first core in anti-aligned models that include the Hall effect is a robust result.  

Concurrent to the study of \citet{Tsukamoto+2015hall}, \citet{WursterPriceBate2016} studied the evolution of the disc until \sm5000~yr after the formation of the protostar; these simulations included sink particles and the barotropic equation of state.  In their aligned models that include, respectively, only the Hall effect and all three non-ideal MHD terms, no discs formed.  In their anti-aligned models, \sm38 and \sm15~au discs formed, respectively.  The disc radius in the non-ideal model with the three terms remained approximately constant until the end of the simulation, while the model with only the Hall effect decreased slightly to \sm20~au by 5000~yr.  Neither disc formed a bar-mode instability, since this instability would be stabilized by the sink particle.  The disc in the non-ideal model is smaller than the \sm25~au discs from the previous studies, but both sink particles and the barotropic equation of state have been shown to reduce the disc sizes compared to models that exclude sink particles and include radiation hydrodynamics \citep[e.g.][]{Tomida+2013,MachidaInutsukaMatsumoto2014,LewisBate2018,WursterBatePrice2018hd}.

Finally, the effect of the cosmic ray ionization rate was studied in \citet{WursterBatePrice2018sd}.  When the ionization rate is increased by even a factor of ten compared to the canonical value for the interstellar medium, the angular momentum in the first core is low enough such that no discs form, independent of magnetic field orientation.  This indicates the importance of the value of the cosmic ray ionization rate: if it is too high, then rotationally supported discs cannot form, despite the initial conditions.  This may have implications for, e.g., starburst galaxies  \citep[e.g.][]{BisbasPapadopoulosViti2015} or the Galactic Centre  \citep[e.g.][]{Oka+2005} where the cosmic ray ionization rate is 10-100 times higher than the canonical value of $\zeta_\text{cr,0} = 10^{-17}$ s$^{-1}$.

%------------------------------------------------------------------------------------------------------------------------------------------------------------------------------------------------------------------------------------------------
\section{Conclusion}
\label{sec:conc}

Protostellar discs have been observed around young stars at all stages of evolution, and naturally form alongside their host star.  The star forming environment has strong magnetic fields, low ionization fractions, and non-laminar velocity flows. The focus of this article is on how the magnetic field affects the formation of disc in lightly ionized molecular cloud cores where non-ideal MHD effects are important. 

Early numerical simulations of star formation that include strong magnetic fields under the ideal MHD approximation fail to form discs during the star forming process; this is known as the magnetic braking catastrophe.  Given that discs are observed, this highlighted the need for including additional physical ingredients in the simulations.  Several studies continued using the ideal MHD approximation, but made their initial conditions less idealized.  By misaligning the magnetic field and rotation vectors, larger or smaller discs could form depending on the study, hence likely depending on the initial conditions.  The misalignment can be naturally induced by a turbulent velocity field, which can speed up the removal of magnetic flux near the centre and thus promote disc formation through enhanced magnetic reconnection; however, increasing turbulence has also been numerically observed to hinder disc formation.  Given that reconnection in ideal MHD simulations is necessarily realised numerically, this process needs to be evaluated more carefully.  

Molecular clouds are observed to be mostly neutral, and the interaction between neutral and charged particles gives rise to three non-ideal effects: Ohmic resistivity, ambipolar diffusion and the Hall effect.  When ambipolar diffusion is included in numerical simulations, small discs of 1-5~au are expected to form over an extended period of time; even larger discs can form if very small grains are removed from the grain size distribution.  When the Hall effect is included and the magnetic field vector is anti-aligned with the rotation axis, larger discs of 25~au or more form.   However, these results may be subject to initial conditions and the choice of parameters such as dust grain properties or ionisation rates.

In summary, considerable progress has been made in averting the magnetic braking catastrophe, through turbulence and related field-rotation misalignment in the ideal MHD limit, enhanced ambipolar diffusion, and especially the Hall effect in the case of anti-aligned magnetic field and rotation axis. Further progress is expected when these and other effects are taken into account together, especially in simulations that can run to the end of the protostellar accretion phase of star formation. These more comprehensive models will be guided by, and be used to interpret, the increasingly more detailed multi-scale observations of the gas kinematics and magnetic fields in the ALMA era.

\section*{Acknowledgments}
We would like to thank the anonymous referees for very thorough and insightful reports which have greatly contributed to the quality of this manuscript.
JW acknowledges support from the European Research Council under the European Community's Seventh Framework Programme (FP7/2007- 2013 grant agreement no. 339248).  Z-Y L is supported in part by NASA NNX14AB38G, NASA 80NSSC18K1095 and NSF AST-1815784 and AST-1716259.  The calculations for this review were performed on the DiRAC Complexity machine, jointly funded by STFC and the Large Facilities Capital Fund of BIS  (STFC grants ST/K000373/1, ST/K0003259/1 and ST/M006948/1), and the University of Exeter Supercomputer, a DiRAC Facility jointly funded by STFC, the Large Facilities Capital Fund of BIS, and the University of Exeter%Isca and Complexity
, and the RIVANNA computer cluster at the University of Virginia.
%------------------------------------------------------------------------------------------------------------------------------------------------------------------------------------------------------------------------------------------------
\bibliographystyle{frontiersinSCNS_ENG_HUMS} \bibliography{frontiers.bib}

\begin{thebibliography}{191}
\providecommand{\natexlab}[1]{#1}
\expandafter\ifx\csname urlstyle\endcsname\relax
  \providecommand{\doi}[1]{doi:\discretionary{}{}{}#1}\else
  \providecommand{\doi}{doi:\discretionary{}{}{}\begingroup
  \urlstyle{rm}\Url}\fi
\providecommand{\selectlanguage}[1]{\relax}
\providecommand{\bibAnnoteFile}[1]{%
  \IfFileExists{#1}{\begin{quotation}\noindent\textsc{Key:} #1\\
  \textsc{Annotation:}\ \input{#1}\end{quotation}}{}}
\providecommand{\bibAnnote}[2]{%
  \begin{quotation}\noindent\textsc{Key:} #1\\
  \textsc{Annotation:}\ #2\end{quotation}}

\bibitem[{{Allen} et~al.(2003){Allen}, {Li}, and {Shu}}]{AllenLiShu2003}
{Allen}, A., {Li}, Z.-Y., and {Shu}, F.~H. (2003).
\newblock {Collapse of Magnetized Singular Isothermal Toroids. II. Rotation and
  Magnetic Braking}.
\newblock \emph{\apj} 599, 363--379.
\newblock \doi{10.1086/379243}
\bibAnnoteFile{AllenLiShu2003}

\bibitem[{{Alves} et~al.(2018){Alves}, {Girart}, {Padovani}, {Galli}, {Franco},
  {Caselli} et~al.}]{Alves+2018}
{Alves}, F.~O., {Girart}, J.~M., {Padovani}, M., {Galli}, D., {Franco},
  G.~A.~P., {Caselli}, P., et~al. (2018).
\newblock {Magnetic field in a young circumbinary disk}.
\newblock \emph{\aap} 616, A56.
\newblock \doi{10.1051/0004-6361/201832935}
\bibAnnoteFile{Alves+2018}

\bibitem[{{Andersson} et~al.(2015){Andersson}, {Lazarian}, and
  {Vaillancourt}}]{AnderssonLazarianVaillancourt2015}
{Andersson}, B.-G., {Lazarian}, A., and {Vaillancourt}, J.~E. (2015).
\newblock {Interstellar Dust Grain Alignment}.
\newblock \emph{\araa} 53, 501--539.
\newblock \doi{10.1146/annurev-astro-082214-122414}
\bibAnnoteFile{AnderssonLazarianVaillancourt2015}

\bibitem[{{Babcock} and {Cowling}(1953)}]{BabcockCowling1953}
{Babcock}, H.~W. and {Cowling}, T.~G. (1953).
\newblock {General magnetic fields in the Sun and stars (Report on progress of
  astronomy)}.
\newblock \emph{\mnras} 113, 357--381.
\newblock \doi{10.1093/mnras/113.3.357}
\bibAnnoteFile{BabcockCowling1953}

\bibitem[{{Bacciotti} et~al.(2018){Bacciotti}, {Girart}, {Padovani}, {Podio},
  {Paladino}, {Testi} et~al.}]{Bacciotti+2018}
{Bacciotti}, F., {Girart}, J.~M., {Padovani}, M., {Podio}, L., {Paladino}, R.,
  {Testi}, L., et~al. (2018).
\newblock {ALMA Observations of Polarized Emission toward the CW Tau and DG Tau
  Protoplanetary Disks: Constraints on Dust Grain Growth and Settling}.
\newblock \emph{\apjl} 865, L12.
\newblock \doi{10.3847/2041-8213/aadf87}
\bibAnnoteFile{Bacciotti+2018}

\bibitem[{{Bai}(2014)}]{Bai2014}
{Bai}, X.-N. (2014).
\newblock {Hall-effect-Controlled Gas Dynamics in Protoplanetary Disks. I. Wind
  Solutions at the Inner Disk}.
\newblock \emph{\apj} 791, 137.
\newblock \doi{10.1088/0004-637X/791/2/137}
\bibAnnoteFile{Bai2014}

\bibitem[{{Basu} and {Mouschovias}(1994)}]{BasuMouschovias1994}
{Basu}, S. and {Mouschovias}, T.~C. (1994).
\newblock {Magnetic braking, ambipolar diffusion, and the formation of cloud
  cores and protostars. 1: Axisymmetric solutions}.
\newblock \emph{\apj} 432, 720--741.
\newblock \doi{10.1086/174611}
\bibAnnoteFile{BasuMouschovias1994}

\bibitem[{{Basu} and {Mouschovias}(1995)}]{BasuMouschovias1995b}
{Basu}, S. and {Mouschovias}, T.~C. (1995).
\newblock {Magnetic Braking, Ambipolar Diffusion, and the Formation of Cloud
  Cores and Protostars. III. Effect of the Initial Mass-to-Flux Ratio}.
\newblock \emph{\apj} 453, 271.
\newblock \doi{10.1086/176387}
\bibAnnoteFile{BasuMouschovias1995b}

\bibitem[{{Bate}(1998)}]{Bate1998}
{Bate}, M.~R. (1998).
\newblock {Collapse of a Molecular Cloud Core to Stellar Densities: The First
  Three-dimensional Calculations}.
\newblock \emph{\apjl} 508, L95--L98.
\newblock \doi{10.1086/311719}
\bibAnnoteFile{Bate1998}

\bibitem[{{Bate}(2010)}]{Bate2010}
{Bate}, M.~R. (2010).
\newblock {Collapse of a molecular cloud core to stellar densities: the
  radiative impact of stellar core formation on the circumstellar disc}.
\newblock \emph{\mnras} 404, L79--L83.
\newblock \doi{10.1111/j.1745-3933.2010.00839.x}
\bibAnnoteFile{Bate2010}

\bibitem[{{Bate}(2011)}]{Bate2011}
{Bate}, M.~R. (2011).
\newblock {Collapse of a molecular cloud core to stellar densities: the
  formation and evolution of pre-stellar discs}.
\newblock \emph{\mnras} 417, 2036--2056.
\newblock \doi{10.1111/j.1365-2966.2011.19386.x}
\bibAnnoteFile{Bate2011}

\bibitem[{{Bate} et~al.(2014){Bate}, {Tricco}, and
  {Price}}]{BateTriccoPrice2014}
{Bate}, M.~R., {Tricco}, T.~S., and {Price}, D.~J. (2014).
\newblock {Collapse of a molecular cloud core to stellar densities:
  stellar-core and outflow formation in radiation magnetohydrodynamic
  simulations}.
\newblock \emph{\mnras} 437, 77--95.
\newblock \doi{10.1093/mnras/stt1865}
\bibAnnoteFile{BateTriccoPrice2014}

\bibitem[{{Bergin} and {Tafalla}(2007)}]{BerginTafalla2007}
{Bergin}, E.~A. and {Tafalla}, M. (2007).
\newblock {Cold Dark Clouds: The Initial Conditions for Star Formation}.
\newblock \emph{\araa} 45, 339--396.
\newblock \doi{10.1146/annurev.astro.45.071206.100404}
\bibAnnoteFile{BerginTafalla2007}

\bibitem[{{Bisbas} et~al.(2015){Bisbas}, {Papadopoulos}, and
  {Viti}}]{BisbasPapadopoulosViti2015}
{Bisbas}, T.~G., {Papadopoulos}, P.~P., and {Viti}, S. (2015).
\newblock {Effective Destruction of CO by Cosmic Rays: Implications for Tracing
  H$_{2}$ Gas in the Universe}.
\newblock \emph{\apj} 803, 37.
\newblock \doi{10.1088/0004-637X/803/1/37}
\bibAnnoteFile{BisbasPapadopoulosViti2015}

\bibitem[{{Boss}(1993)}]{Boss1993}
{Boss}, A.~P. (1993).
\newblock {Collapse and fragmentation of molecular cloud cores. I - Moderately
  centrally condensed cores}.
\newblock \emph{\apj} 410, 157--167.
\newblock \doi{10.1086/172734}
\bibAnnoteFile{Boss1993}

\bibitem[{{Boss}(1997)}]{Boss1997}
{Boss}, A.~P. (1997).
\newblock {Giant planet formation by gravitational instability.}
\newblock \emph{Science} 276, 1836--1839.
\newblock \doi{10.1126/science.276.5320.1836}
\bibAnnoteFile{Boss1997}

\bibitem[{{Boss}(1999)}]{Boss1999}
{Boss}, A.~P. (1999).
\newblock {Collapse and Fragmentation of Molecular Cloud Cores. VI. Slowly
  Rotating Magnetic Clouds}.
\newblock \emph{\apj} 520, 744--750.
\newblock \doi{10.1086/307479}
\bibAnnoteFile{Boss1999}

\bibitem[{{Boss}(2002)}]{Boss2002}
{Boss}, A.~P. (2002).
\newblock {Collapse and Fragmentation of Molecular Cloud Cores. VII. Magnetic
  Fields and Multiple Protostar Formation}.
\newblock \emph{\apj} 568, 743--753.
\newblock \doi{10.1086/339040}
\bibAnnoteFile{Boss2002}

\bibitem[{{Boss}(2005)}]{Boss2005}
{Boss}, A.~P. (2005).
\newblock {Collapse and Fragmentation of Molecular Cloud Cores. VIII.
  Magnetically Supported Infinite Sheets}.
\newblock \emph{\apj} 622, 393--403.
\newblock \doi{10.1086/428113}
\bibAnnoteFile{Boss2005}

\bibitem[{{Boss}(2007)}]{Boss2007}
{Boss}, A.~P. (2007).
\newblock {Collapse and Fragmentation of Molecular Cloud Cores. IX. Magnetic
  Braking of Initially Filamentary Clouds}.
\newblock \emph{\apj} 658, 1136--1143.
\newblock \doi{10.1086/512061}
\bibAnnoteFile{Boss2007}

\bibitem[{{Boss}(2009)}]{Boss2009}
{Boss}, A.~P. (2009).
\newblock {Collapse and Fragmentation of Molecular Cloud Cores. X. Magnetic
  Braking of Prolate and Oblate Cores}.
\newblock \emph{\apj} 697, 1940--1945.
\newblock \doi{10.1088/0004-637X/697/2/1940}
\bibAnnoteFile{Boss2009}

\bibitem[{{Boss} and {Myhill}(1995)}]{BossMyhill1995}
{Boss}, A.~P. and {Myhill}, E.~A. (1995).
\newblock {Collapse and Fragmentation of Molecular Cloud Cores. III. Initial
  Differential Rotation}.
\newblock \emph{\apj} 451, 218.
\newblock \doi{10.1086/176213}
\bibAnnoteFile{BossMyhill1995}

\bibitem[{{Bourke} et~al.(2001){Bourke}, {Myers}, {Robinson}, and
  {Hyland}}]{Bourke+2001}
{Bourke}, T.~L., {Myers}, P.~C., {Robinson}, G., and {Hyland}, A.~R. (2001).
\newblock {New OH Zeeman Measurements of Magnetic Field Strengths in Molecular
  Clouds}.
\newblock \emph{\apj} 554, 916--932.
\newblock \doi{10.1086/321405}
\bibAnnoteFile{Bourke+2001}

\bibitem[{{Braiding} and {Wardle}(2012{\natexlab{a}})}]{BraidingWardle2012sf}
{Braiding}, C.~R. and {Wardle}, M. (2012{\natexlab{a}}).
\newblock { The Hall effect in star formation}.
\newblock \emph{\mnras} 422, 261--281.
\newblock \doi{10.1111/j.1365-2966.2012.20601.x}
\bibAnnoteFile{BraidingWardle2012sf}

\bibitem[{{Braiding} and
  {Wardle}(2012{\natexlab{b}})}]{BraidingWardle2012accretion}
{Braiding}, C.~R. and {Wardle}, M. (2012{\natexlab{b}}).
\newblock {The Hall effect in accretion flows}.
\newblock \emph{\mnras} 427, 3188--3195.
\newblock \doi{10.1111/j.1365-2966.2012.22001.x}
\bibAnnoteFile{BraidingWardle2012accretion}

\bibitem[{{Chandran}(2000)}]{Chandran2000}
{Chandran}, B.~D.~G. (2000).
\newblock {Confinement and Isotropization of Galactic Cosmic Rays by
  Molecular-Cloud Magnetic Mirrors When Turbulent Scattering Is Weak}.
\newblock \emph{\apj} 529, 513--535.
\newblock \doi{10.1086/308232}
\bibAnnoteFile{Chandran2000}

\bibitem[{{Choi} et~al.(2009){Choi}, {Kim}, and {Wiita}}]{ChoiKimWiita2009}
{Choi}, E., {Kim}, J., and {Wiita}, P.~J. (2009).
\newblock {An Explicit Scheme for Incorporating Ambipolar Diffusion in a
  Magnetohydrodynamics Code}.
\newblock \emph{\apjs} 181, 413--420.
\newblock \doi{10.1088/0067-0049/181/2/413}
\bibAnnoteFile{ChoiKimWiita2009}

\bibitem[{{Ciolek} and {Mouschovias}(1993)}]{CiolekMouschovias1993}
{Ciolek}, G.~E. and {Mouschovias}, T.~C. (1993).
\newblock {Ambipolar Diffusion, Interstellar Dust, and the Formation of Cloud
  Cores and Protostars. I. Basic Physics and Formulation of the Problem}.
\newblock \emph{\apj} 418, 774.
\newblock \doi{10.1086/173435}
\bibAnnoteFile{CiolekMouschovias1993}

\bibitem[{{Ciolek} and {Mouschovias}(1994)}]{CiolekMouschovias1994}
{Ciolek}, G.~E. and {Mouschovias}, T.~C. (1994).
\newblock {Ambipolar diffusion, interstellar dust, and the formation of cloud
  cores and protostars. 3: Typical axisymmetric solutions}.
\newblock \emph{\apj} 425, 142--160.
\newblock \doi{10.1086/173971}
\bibAnnoteFile{CiolekMouschovias1994}

\bibitem[{{Codella} et~al.(2014){Codella}, {Cabrit}, {Gueth}, {Podio},
  {Leurini}, {Bachiller} et~al.}]{Codella+2014}
{Codella}, C., {Cabrit}, S., {Gueth}, F., {Podio}, L., {Leurini}, S.,
  {Bachiller}, R., et~al. (2014).
\newblock {The ALMA view of the protostellar system HH212. The wind, the
  cavity, and the disk}.
\newblock \emph{\aap} 568, L5.
\newblock \doi{10.1051/0004-6361/201424103}
\bibAnnoteFile{Codella+2014}

\bibitem[{{Commer{\c c}on} et~al.(2010){Commer{\c c}on}, {Hennebelle}, {Audit},
  {Chabrier}, and {Teyssier}}]{Commercon+2010}
{Commer{\c c}on}, B., {Hennebelle}, P., {Audit}, E., {Chabrier}, G., and
  {Teyssier}, R. (2010).
\newblock {Protostellar collapse: radiative and magnetic feedbacks on
  small-scale fragmentation}.
\newblock \emph{\aap} 510, L3.
\newblock \doi{10.1051/0004-6361/200913597}
\bibAnnoteFile{Commercon+2010}

\bibitem[{{Cox} et~al.(2018){Cox}, {Harris}, {Looney}, {Li}, {Yang}, {Tobin}
  et~al.}]{Cox+2018}
{Cox}, E.~G., {Harris}, R.~J., {Looney}, L.~W., {Li}, Z.-Y., {Yang}, H.,
  {Tobin}, J.~J., et~al. (2018).
\newblock {ALMA's Polarized View of 10 Protostars in the Perseus Molecular
  Cloud}.
\newblock \emph{\apj} 855, 92.
\newblock \doi{10.3847/1538-4357/aaacd2}
\bibAnnoteFile{Cox+2018}

\bibitem[{{Cox} et~al.(2015){Cox}, {Harris}, {Looney}, {Segura-Cox}, {Tobin},
  {Li} et~al.}]{Cox+2015}
{Cox}, E.~G., {Harris}, R.~J., {Looney}, L.~W., {Segura-Cox}, D.~M., {Tobin},
  J., {Li}, Z.-Y., et~al. (2015).
\newblock {High-resolution 8 mm and 1 cm Polarization of IRAS 4A from the VLA
  Nascent Disk and Multiplicity (VANDAM) Survey}.
\newblock \emph{\apjl} 814, L28.
\newblock \doi{10.1088/2041-8205/814/2/L28}
\bibAnnoteFile{Cox+2015}

\bibitem[{{Crutcher}(1999)}]{Crutcher1999}
{Crutcher}, R.~M. (1999).
\newblock {Magnetic Fields in Molecular Clouds: Observations Confront Theory}.
\newblock \emph{\apj} 520, 706--713.
\newblock \doi{10.1086/307483}
\bibAnnoteFile{Crutcher1999}

\bibitem[{{Dapp} and {Basu}(2010)}]{DappBasu2010}
{Dapp}, W.~B. and {Basu}, S. (2010).
\newblock {Averting the magnetic braking catastrophe on small scales: disk
  formation due to Ohmic dissipation}.
\newblock \emph{\aap} 521, L56.
\newblock \doi{10.1051/0004-6361/201015700}
\bibAnnoteFile{DappBasu2010}

\bibitem[{{Dapp} et~al.(2012){Dapp}, {Basu}, and {Kunz}}]{DappBasuKunz2012}
{Dapp}, W.~B., {Basu}, S., and {Kunz}, M.~W. (2012).
\newblock {Bridging the gap: disk formation in the Class 0 phase with ambipolar
  diffusion and Ohmic dissipation}.
\newblock \emph{\aap} 541, A35.
\newblock \doi{10.1051/0004-6361/201117876}
\bibAnnoteFile{DappBasuKunz2012}

\bibitem[{{Davis} and {Greenstein}(1951)}]{DavisGreenstein1951}
{Davis}, L., Jr. and {Greenstein}, J.~L. (1951).
\newblock {The Polarization of Starlight by Aligned Dust Grains.}
\newblock \emph{\apj} 114, 206.
\newblock \doi{10.1086/145464}
\bibAnnoteFile{DavisGreenstein1951}

\bibitem[{{Dent} et~al.(2019){Dent}, {Pinte}, {Cortes}, {M{\'e}nard}, {Hales},
  {Fomalont} et~al.}]{Dent+2019}
{Dent}, W.~R.~F., {Pinte}, C., {Cortes}, P.~C., {M{\'e}nard}, F., {Hales}, A.,
  {Fomalont}, E., et~al. (2019).
\newblock {Submillimetre dust polarization and opacity in the HD163296
  protoplanetary ring system}.
\newblock \emph{\mnras} 482, L29--L33.
\newblock \doi{10.1093/mnrasl/sly181}
\bibAnnoteFile{Dent+2019}

\bibitem[{{Draine} and {Lee}(1984)}]{DraineLee1984}
{Draine}, B.~T. and {Lee}, H.~M. (1984).
\newblock {Optical properties of interstellar graphite and silicate grains}.
\newblock \emph{\apj} 285, 89--108.
\newblock \doi{10.1086/162480}
\bibAnnoteFile{DraineLee1984}

\bibitem[{{Duffin} and {Pudritz}(2009)}]{DuffinPudritz2009}
{Duffin}, D.~F. and {Pudritz}, R.~E. (2009).
\newblock {The Early History of Protostellar Disks, Outflows, and Binary
  Stars}.
\newblock \emph{\apjl} 706, L46--L51.
\newblock \doi{10.1088/0004-637X/706/1/L46}
\bibAnnoteFile{DuffinPudritz2009}

\bibitem[{{Fielding} et~al.(2015){Fielding}, {McKee}, {Socrates}, {Cunningham},
  and {Klein}}]{Fielding+2015}
{Fielding}, D.~B., {McKee}, C.~F., {Socrates}, A., {Cunningham}, A.~J., and
  {Klein}, R.~I. (2015).
\newblock {The turbulent origin of spin-orbit misalignment in planetary
  systems}.
\newblock \emph{\mnras} 450, 3306--3318.
\newblock \doi{10.1093/mnras/stv836}
\bibAnnoteFile{Fielding+2015}

\bibitem[{{Galli} et~al.(2006){Galli}, {Lizano}, {Shu}, and
  {Allen}}]{Galli+2006}
{Galli}, D., {Lizano}, S., {Shu}, F.~H., and {Allen}, A. (2006).
\newblock {Gravitational Collapse of Magnetized Clouds. I. Ideal
  Magnetohydrodynamic Accretion Flow}.
\newblock \emph{\apj} 647, 374--381.
\newblock \doi{10.1086/505257}
\bibAnnoteFile{Galli+2006}

\bibitem[{{Galli} and {Shu}(1993)}]{GalliShu1993}
{Galli}, D. and {Shu}, F.~H. (1993).
\newblock {Collapse of Magnetized Molecular Cloud Cores. II. Numerical
  Results}.
\newblock \emph{\apj} 417, 243.
\newblock \doi{10.1086/173306}
\bibAnnoteFile{GalliShu1993}

\bibitem[{{Girart} et~al.(2018){Girart}, {Fern{\'a}ndez-L{\'o}pez}, {Li},
  {Yang}, {Estalella}, {Anglada} et~al.}]{Girart+2018}
{Girart}, J.~M., {Fern{\'a}ndez-L{\'o}pez}, M., {Li}, Z.-Y., {Yang}, H.,
  {Estalella}, R., {Anglada}, G., et~al. (2018).
\newblock {Resolving the Polarized Dust Emission of the Disk around the Massive
  Star Powering the HH 80-81 Radio Jet}.
\newblock \emph{\apjl} 856, L27.
\newblock \doi{10.3847/2041-8213/aab76b}
\bibAnnoteFile{Girart+2018}

\bibitem[{{Goldsmith} and {Arquilla}(1985)}]{GoldsmithArquilla1985}
{Goldsmith}, P.~F. and {Arquilla}, R. (1985).
\newblock {Rotation in dark clouds}.
\newblock In \emph{Protostars and Planets II}, eds. D.~C. {Black} and M.~S.
  {Matthews}. 137--149
\bibAnnoteFile{GoldsmithArquilla1985}

\bibitem[{{Goodman} et~al.(1993){Goodman}, {Benson}, {Fuller}, and
  {Myers}}]{Goodman+1993}
{Goodman}, A.~A., {Benson}, P.~J., {Fuller}, G.~A., and {Myers}, P.~C. (1993).
\newblock {Dense cores in dark clouds. VIII - Velocity gradients}.
\newblock \emph{\apj} 406, 528--547.
\newblock \doi{10.1086/172465}
\bibAnnoteFile{Goodman+1993}

\bibitem[{{Gray} et~al.(2018){Gray}, {McKee}, and {Klein}}]{GrayMcKeeKlein2018}
{Gray}, W.~J., {McKee}, C.~F., and {Klein}, R.~I. (2018).
\newblock {Effect of angular momentum alignment and strong magnetic fields on
  the formation of protostellar discs}.
\newblock \emph{\mnras} 473, 2124--2143.
\newblock \doi{10.1093/mnras/stx2406}
\bibAnnoteFile{GrayMcKeeKlein2018}

\bibitem[{{Hall}(1949)}]{Hall1949}
{Hall}, J.~S. (1949).
\newblock {Observations of the Polarized Light from Stars}.
\newblock \emph{Science} 109, 166--167.
\newblock \doi{10.1126/science.109.2825.166}
\bibAnnoteFile{Hall1949}

\bibitem[{{Harris} et~al.(2018){Harris}, {Cox}, {Looney}, {Li}, {Yang},
  {Fern{\'a}ndez-L{\'o}pez} et~al.}]{Harris+2018}
{Harris}, R.~J., {Cox}, E.~G., {Looney}, L.~W., {Li}, Z.-Y., {Yang}, H.,
  {Fern{\'a}ndez-L{\'o}pez}, M., et~al. (2018).
\newblock {ALMA Observations of Polarized 872 {$\mu$}m Dust Emission from the
  Protostellar Systems VLA 1623 and L1527}.
\newblock \emph{\apj} 861, 91.
\newblock \doi{10.3847/1538-4357/aac6ec}
\bibAnnoteFile{Harris+2018}

\bibitem[{{Heiles} and {Crutcher}(2005)}]{HeilesCrutcher2005}
{Heiles}, C. and {Crutcher}, R. (2005).
\newblock {Magnetic Fields in Diffuse HI and Molecular Clouds}.
\newblock In \emph{Cosmic Magnetic Fields}, eds. R.~{Wielebinski} and
  R.~{Beck}. vol. 664 of \emph{Lecture Notes in Physics, Berlin Springer
  Verlag}, 137.
\newblock \doi{10.1007/11369875_7}
\bibAnnoteFile{HeilesCrutcher2005}

\bibitem[{{Hennebelle} and {Ciardi}(2009)}]{HennebelleCiardi2009}
{Hennebelle}, P. and {Ciardi}, A. (2009).
\newblock {Disk formation during collapse of magnetized protostellar cores}.
\newblock \emph{\aap} 506, L29--L32.
\newblock \doi{10.1051/0004-6361/200913008}
\bibAnnoteFile{HennebelleCiardi2009}

\bibitem[{{Hennebelle} et~al.(2016){Hennebelle}, {Commer{\c c}on}, {Chabrier},
  and {Marchand}}]{Hennebelle+2016}
{Hennebelle}, P., {Commer{\c c}on}, B., {Chabrier}, G., and {Marchand}, P.
  (2016).
\newblock {Magnetically Self-regulated Formation of Early Protoplanetary
  Disks}.
\newblock \emph{\apjl} 830, L8.
\newblock \doi{10.3847/2041-8205/830/1/L8}
\bibAnnoteFile{Hennebelle+2016}

\bibitem[{{Hennebelle} and {Fromang}(2008)}]{HennebelleFromang2008}
{Hennebelle}, P. and {Fromang}, S. (2008).
\newblock {Magnetic processes in a collapsing dense core. I. Accretion and
  ejection}.
\newblock \emph{\aap} 477, 9--24.
\newblock \doi{10.1051/0004-6361:20078309}
\bibAnnoteFile{HennebelleFromang2008}

\bibitem[{{Heyer} and {Brunt}(2004)}]{HeyerBrunt2004}
{Heyer}, M.~H. and {Brunt}, C.~M. (2004).
\newblock {The Universality of Turbulence in Galactic Molecular Clouds}.
\newblock \emph{\apjl} 615, L45--L48.
\newblock \doi{10.1086/425978}
\bibAnnoteFile{HeyerBrunt2004}

\bibitem[{{Higuchi} et~al.(2018){Higuchi}, {Machida}, and
  {Susa}}]{HiguchiMachidaSusa2018}
{Higuchi}, K., {Machida}, M.~N., and {Susa}, H. (2018).
\newblock {Evolution of magnetic fields in collapsing star-forming clouds under
  different environments}.
\newblock \emph{\mnras} 475, 3331--3347.
\newblock \doi{10.1093/mnras/sty046}
\bibAnnoteFile{HiguchiMachidaSusa2018}

\bibitem[{{Hiltner}(1949)}]{Hiltner1949}
{Hiltner}, W.~A. (1949).
\newblock {On the Presence of Polarization in the Continuous Radiation of
  Stars. II.}
\newblock \emph{\apj} 109, 471.
\newblock \doi{10.1086/145151}
\bibAnnoteFile{Hiltner1949}

\bibitem[{{Hull} et~al.(2013){Hull}, {Plambeck}, {Bolatto}, {Bower},
  {Carpenter}, {Crutcher} et~al.}]{Hull+2013}
{Hull}, C.~L.~H., {Plambeck}, R.~L., {Bolatto}, A.~D., {Bower}, G.~C.,
  {Carpenter}, J.~M., {Crutcher}, R.~M., et~al. (2013).
\newblock {Misalignment of Magnetic Fields and Outflows in Protostellar Cores}.
\newblock \emph{\apj} 768, 159.
\newblock \doi{10.1088/0004-637X/768/2/159}
\bibAnnoteFile{Hull+2013}

\bibitem[{{Hull} et~al.(2018){Hull}, {Yang}, {Li}, {Kataoka}, {Stephens},
  {Andrews} et~al.}]{Hull+2018}
{Hull}, C.~L.~H., {Yang}, H., {Li}, Z.-Y., {Kataoka}, A., {Stephens}, I.~W.,
  {Andrews}, S., et~al. (2018).
\newblock {ALMA Observations of Polarization from Dust Scattering in the IM Lup
  Protoplanetary Disk}.
\newblock \emph{\apj} 860, 82.
\newblock \doi{10.3847/1538-4357/aabfeb}
\bibAnnoteFile{Hull+2018}

\bibitem[{{Igea} and {Glassgold}(1999)}]{IgeaGlassgold1999}
{Igea}, J. and {Glassgold}, A.~E. (1999).
\newblock {X-Ray Ionization of the Disks of Young Stellar Objects}.
\newblock \emph{\apj} 518, 848--858.
\newblock \doi{10.1086/307302}
\bibAnnoteFile{IgeaGlassgold1999}

\bibitem[{{Inoue} and {Inutsuka}(2008)}]{InoueInutsuka2008}
{Inoue}, T. and {Inutsuka}, S.-i. (2008).
\newblock {Two-Fluid Magnetohydrodynamic Simulations of Converging H I Flows in
  the Interstellar Medium. I. Methodology and Basic Results}.
\newblock \emph{\apj} 687, 303-310.
\newblock \doi{10.1086/590528}
\bibAnnoteFile{InoueInutsuka2008}

\bibitem[{{Inoue} and {Inutsuka}(2009)}]{InoueInutsuka2009}
{Inoue}, T. and {Inutsuka}, S.-i. (2009).
\newblock {Two-Fluid Magnetohydrodynamics Simulations of Converging H I Flows
  in the Interstellar Medium. II. Are Molecular Clouds Generated Directly from
  a Warm Neutral Medium?}
\newblock \emph{\apj} 704, 161--169.
\newblock \doi{10.1088/0004-637X/704/1/161}
\bibAnnoteFile{InoueInutsuka2009}

\bibitem[{{Inoue} et~al.(2007){Inoue}, {Inutsuka}, and
  {Koyama}}]{InoueInutsukaKoyama2007}
{Inoue}, T., {Inutsuka}, S.-i., and {Koyama}, H. (2007).
\newblock {The Role of Ambipolar Diffusion in the Formation Process of
  Moderately Magnetized Diffuse Clouds}.
\newblock \emph{\apjl} 658, L99--L102.
\newblock \doi{10.1086/514816}
\bibAnnoteFile{InoueInutsukaKoyama2007}

\bibitem[{{Inutsuka} et~al.(2010){Inutsuka}, {Machida}, and
  {Matsumoto}}]{InutsukaMachidaMatsumoto2010}
{Inutsuka}, S.-i., {Machida}, M.~N., and {Matsumoto}, T. (2010).
\newblock {Emergence of Protoplanetary Disks and Successive Formation of
  Gaseous Planets by Gravitational Instability}.
\newblock \emph{\apjl} 718, L58--L62.
\newblock \doi{10.1088/2041-8205/718/2/L58}
\bibAnnoteFile{InutsukaMachidaMatsumoto2010}

\bibitem[{{Joos} et~al.(2012){Joos}, {Hennebelle}, and
  {Ciardi}}]{JoosHennebelleCiardi2012}
{Joos}, M., {Hennebelle}, P., and {Ciardi}, A. (2012).
\newblock {Protostellar disk formation and transport of angular momentum during
  magnetized core collapse}.
\newblock \emph{\aap} 543, A128.
\newblock \doi{10.1051/0004-6361/201118730}
\bibAnnoteFile{JoosHennebelleCiardi2012}

\bibitem[{{Joos} et~al.(2013){Joos}, {Hennebelle}, {Ciardi}, and
  {Fromang}}]{Joos+2013}
{Joos}, M., {Hennebelle}, P., {Ciardi}, A., and {Fromang}, S. (2013).
\newblock {The influence of turbulence during magnetized core collapse and its
  consequences on low-mass star formation}.
\newblock \emph{\aap} 554, A17.
\newblock \doi{10.1051/0004-6361/201220649}
\bibAnnoteFile{Joos+2013}

\bibitem[{{Kataoka} et~al.(2015){Kataoka}, {Muto}, {Momose}, {Tsukagoshi},
  {Fukagawa}, {Shibai} et~al.}]{Kataoka+2015}
{Kataoka}, A., {Muto}, T., {Momose}, M., {Tsukagoshi}, T., {Fukagawa}, M.,
  {Shibai}, H., et~al. (2015).
\newblock {Millimeter-wave Polarization of Protoplanetary Disks due to Dust
  Scattering}.
\newblock \emph{\apj} 809, 78.
\newblock \doi{10.1088/0004-637X/809/1/78}
\bibAnnoteFile{Kataoka+2015}

\bibitem[{{Kataoka} et~al.(2016){Kataoka}, {Tsukagoshi}, {Momose}, {Nagai},
  {Muto}, {Dullemond} et~al.}]{Kataoka+2016polarisation}
{Kataoka}, A., {Tsukagoshi}, T., {Momose}, M., {Nagai}, H., {Muto}, T.,
  {Dullemond}, C.~P., et~al. (2016).
\newblock {Submillimeter Polarization Observation of the Protoplanetary Disk
  around HD 142527}.
\newblock \emph{\apjl} 831, L12.
\newblock \doi{10.3847/2041-8205/831/2/L12}
\bibAnnoteFile{Kataoka+2016polarisation}

\bibitem[{{Kataoka} et~al.(2017){Kataoka}, {Tsukagoshi}, {Pohl}, {Muto},
  {Nagai}, {Stephens} et~al.}]{Kataoka+2017}
{Kataoka}, A., {Tsukagoshi}, T., {Pohl}, A., {Muto}, T., {Nagai}, H.,
  {Stephens}, I.~W., et~al. (2017).
\newblock {The Evidence of Radio Polarization Induced by the Radiative Grain
  Alignment and Self-scattering of Dust Grains in a Protoplanetary Disk}.
\newblock \emph{\apjl} 844, L5.
\newblock \doi{10.3847/2041-8213/aa7e33}
\bibAnnoteFile{Kataoka+2017}

\bibitem[{{Krasnopolsky} et~al.(2010){Krasnopolsky}, {Li}, and
  {Shang}}]{KrasnopolskyLiShang2010}
{Krasnopolsky}, R., {Li}, Z.-Y., and {Shang}, H. (2010).
\newblock {Disk Formation Enabled by Enhanced Resistivity}.
\newblock \emph{\apj} 716, 1541--1550.
\newblock \doi{10.1088/0004-637X/716/2/1541}
\bibAnnoteFile{KrasnopolskyLiShang2010}

\bibitem[{{Krasnopolsky} et~al.(2011){Krasnopolsky}, {Li}, and
  {Shang}}]{KrasnopolskyLiShang2011}
{Krasnopolsky}, R., {Li}, Z.-Y., and {Shang}, H. (2011).
\newblock {Disk Formation in Magnetized Clouds Enabled by the Hall Effect}.
\newblock \emph{\apj} 733, 54.
\newblock \doi{10.1088/0004-637X/733/1/54}
\bibAnnoteFile{KrasnopolskyLiShang2011}

\bibitem[{{Krumholz} et~al.(2013){Krumholz}, {Crutcher}, and
  {Hull}}]{KrumholzCrutcherHull2013}
{Krumholz}, M.~R., {Crutcher}, R.~M., and {Hull}, C.~L.~H. (2013).
\newblock {Protostellar Disk Formation Enabled by Weak, Misaligned Magnetic
  Fields}.
\newblock \emph{\apjl} 767, L11.
\newblock \doi{10.1088/2041-8205/767/1/L11}
\bibAnnoteFile{KrumholzCrutcherHull2013}

\bibitem[{{Kunz} and {Mouschovias}(2009)}]{KunzMouschovias2009}
{Kunz}, M.~W. and {Mouschovias}, T.~C. (2009).
\newblock {The Nonisothermal Stage of Magnetic Star Formation. I. Formulation
  of the Problem and Method of Solution}.
\newblock \emph{\apj} 693, 1895--1911.
\newblock \doi{10.1088/0004-637X/693/2/1895}
\bibAnnoteFile{KunzMouschovias2009}

\bibitem[{{Kunz} and {Mouschovias}(2010)}]{KunzMouschovias2010}
{Kunz}, M.~W. and {Mouschovias}, T.~C. (2010).
\newblock {The non-isothermal stage of magnetic star formation - II. Results}.
\newblock \emph{\mnras} 408, 322--341.
\newblock \doi{10.1111/j.1365-2966.2010.17110.x}
\bibAnnoteFile{KunzMouschovias2010}

\bibitem[{{Larson}(1969)}]{Larson1969}
{Larson}, R.~B. (1969).
\newblock {Numerical calculations of the dynamics of collapsing proto-star}.
\newblock \emph{\mnras} 145, 271
\bibAnnoteFile{Larson1969}

\bibitem[{{Lee} et~al.(2018){Lee}, {Li}, {Ching}, {Lai}, and {Yang}}]{Lee+2018}
{Lee}, C.-F., {Li}, Z.-Y., {Ching}, T.-C., {Lai}, S.-P., and {Yang}, H. (2018).
\newblock {ALMA Dust Polarization Observations of Two Young Edge-on
  Protostellar Disks}.
\newblock \emph{\apj} 854, 56.
\newblock \doi{10.3847/1538-4357/aaa769}
\bibAnnoteFile{Lee+2018}

\bibitem[{{Lee} et~al.(2017){Lee}, {Li}, {Ho}, {Hirano}, {Zhang}, and
  {Shang}}]{Lee+2017}
{Lee}, C.-F., {Li}, Z.-Y., {Ho}, P.~T.~P., {Hirano}, N., {Zhang}, Q., and
  {Shang}, H. (2017).
\newblock {First detection of equatorial dark dust lane in a protostellar disk
  at submillimeter wavelength}.
\newblock \emph{Science Advances} 3, e1602935.
\newblock \doi{10.1126/sciadv.1602935}
\bibAnnoteFile{Lee+2017}

\bibitem[{{Lewis} and {Bate}(2017)}]{LewisBate2017}
{Lewis}, B.~T. and {Bate}, M.~R. (2017).
\newblock {The dependence of protostar formation on the geometry and strength
  of the initial magnetic field}.
\newblock \emph{\mnras} 467, 3324--3337.
\newblock \doi{10.1093/mnras/stx271}
\bibAnnoteFile{LewisBate2017}

\bibitem[{{Lewis} and {Bate}(2018)}]{LewisBate2018}
{Lewis}, B.~T. and {Bate}, M.~R. (2018).
\newblock {Shaken and stirred: the effects of turbulence and rotation on disc
  and outflow formation during the collapse of magnetized molecular cloud
  cores}.
\newblock \emph{\mnras} 477, 4241--4256.
\newblock \doi{10.1093/mnras/sty829}
\bibAnnoteFile{LewisBate2018}

\bibitem[{{Lewis} et~al.(2015){Lewis}, {Bate}, and
  {Price}}]{LewisBatePrice2015}
{Lewis}, B.~T., {Bate}, M.~R., and {Price}, D.~J. (2015).
\newblock {Smoothed particle magnetohydrodynamic simulations of protostellar
  outflows with misaligned magnetic field and rotation axes}.
\newblock \emph{\mnras} 451, 288--299.
\newblock \doi{10.1093/mnras/stv957}
\bibAnnoteFile{LewisBatePrice2015}

\bibitem[{{Li} et~al.(2013{\natexlab{a}}){Li}, {Fang}, {Henning}, and
  {Kainulainen}}]{LiFangHenningKainulainen2013}
{Li}, H.-b., {Fang}, M., {Henning}, T., and {Kainulainen}, J.
  (2013{\natexlab{a}}).
\newblock {The link between magnetic fields and filamentary clouds: bimodal
  cloud orientations in the Gould Belt}.
\newblock \emph{\mnras} 436, 3707--3719.
\newblock \doi{10.1093/mnras/stt1849}
\bibAnnoteFile{LiFangHenningKainulainen2013}

\bibitem[{{Li} et~al.(2011){Li}, {Krasnopolsky}, and
  {Shang}}]{LiKrasnopolskyShang2011}
{Li}, Z.-Y., {Krasnopolsky}, R., and {Shang}, H. (2011).
\newblock {Non-ideal MHD Effects and Magnetic Braking Catastrophe in
  Protostellar Disk Formation}.
\newblock \emph{\apj} 738, 180.
\newblock \doi{10.1088/0004-637X/738/2/180}
\bibAnnoteFile{LiKrasnopolskyShang2011}

\bibitem[{{Li} et~al.(2013{\natexlab{b}}){Li}, {Krasnopolsky}, and
  {Shang}}]{LiKrasnopolskyShang2013}
{Li}, Z.-Y., {Krasnopolsky}, R., and {Shang}, H. (2013{\natexlab{b}}).
\newblock {Does Magnetic-field-Rotation Misalignment Solve the Magnetic Braking
  Catastrophe in Protostellar Disk Formation?}
\newblock \emph{\apj} 774, 82.
\newblock \doi{10.1088/0004-637X/774/1/82}
\bibAnnoteFile{LiKrasnopolskyShang2013}

\bibitem[{{Li} et~al.(2014){Li}, {Krasnopolsky}, {Shang}, and {Zhao}}]{Li+2014}
{Li}, Z.-Y., {Krasnopolsky}, R., {Shang}, H., and {Zhao}, B. (2014).
\newblock {On the Role of Pseudodisk Warping and Reconnection in Protostellar
  Disk Formation in Turbulent Magnetized Cores}.
\newblock \emph{\apj} 793, 130.
\newblock \doi{10.1088/0004-637X/793/2/130}
\bibAnnoteFile{Li+2014}

\bibitem[{{Li} and {McKee}(1996)}]{LiMckee1996}
{Li}, Z.-Y. and {McKee}, C.~F. (1996).
\newblock {Hydromagnetic Accretion Shocks around Low-Mass Protostars}.
\newblock \emph{\apj} 464, 373.
\newblock \doi{10.1086/177329}
\bibAnnoteFile{LiMckee1996}

\bibitem[{{Li} and {Shu}(1996)}]{LiShu1996}
{Li}, Z.-Y. and {Shu}, F.~H. (1996).
\newblock {Magnetized Singular Isothermal Toroids}.
\newblock \emph{\apj} 472, 211.
\newblock \doi{10.1086/178056}
\bibAnnoteFile{LiShu1996}

\bibitem[{{Liu} et~al.(2016){Liu}, {Lai}, {Hasegawa}, {Hirano}, {Rao}, {Li}
  et~al.}]{Liu+2016}
{Liu}, H.~B., {Lai}, S.-P., {Hasegawa}, Y., {Hirano}, N., {Rao}, R., {Li},
  I.-H., et~al. (2016).
\newblock {Detection of Linearly Polarized 6.9 mm Continuum Emission from the
  Class 0 Young Stellar Object NGC 1333 IRAS4A}.
\newblock \emph{\apj} 821, 41.
\newblock \doi{10.3847/0004-637X/821/1/41}
\bibAnnoteFile{Liu+2016}

\bibitem[{{Mac Low} et~al.(1995){Mac Low}, {Norman}, {Konigl}, and
  {Wardle}}]{Maclow+1995}
{Mac Low}, M.-M., {Norman}, M.~L., {Konigl}, A., and {Wardle}, M. (1995).
\newblock {Incorporation of ambipolar diffusion into the ZEUS
  magnetohydrodynamics code}.
\newblock \emph{\apj} 442, 726--735.
\newblock \doi{10.1086/175477}
\bibAnnoteFile{Maclow+1995}

\bibitem[{{Machida} and {Hosokawa}(2013)}]{MachidaHosokawa2013}
{Machida}, M.~N. and {Hosokawa}, T. (2013).
\newblock {Evolution of protostellar outflow around low-mass protostar}.
\newblock \emph{\mnras} 431, 1719--1744.
\newblock \doi{10.1093/mnras/stt291}
\bibAnnoteFile{MachidaHosokawa2013}

\bibitem[{{Machida} et~al.(2008){Machida}, {Inutsuka}, and
  {Matsumoto}}]{MachidaInutsukaMatsumoto2008}
{Machida}, M.~N., {Inutsuka}, S.-i., and {Matsumoto}, T. (2008).
\newblock {High- and Low-Velocity Magnetized Outflows in the Star Formation
  Process in a Gravitationally Collapsing Cloud}.
\newblock \emph{\apj} 676, 1088--1108.
\newblock \doi{10.1086/528364}
\bibAnnoteFile{MachidaInutsukaMatsumoto2008}

\bibitem[{{Machida} et~al.(2010){Machida}, {Inutsuka}, and
  {Matsumoto}}]{MachidaInutsukaMatsumoto2010}
{Machida}, M.~N., {Inutsuka}, S.-i., and {Matsumoto}, T. (2010).
\newblock {Formation Process of the Circumstellar Disk: Long-term Simulations
  in the Main Accretion Phase of Star Formation}.
\newblock \emph{\apj} 724, 1006--1020.
\newblock \doi{10.1088/0004-637X/724/2/1006}
\bibAnnoteFile{MachidaInutsukaMatsumoto2010}

\bibitem[{{Machida} et~al.(2011){Machida}, {Inutsuka}, and
  {Matsumoto}}]{MachidaInutsukaMatsumoto2011}
{Machida}, M.~N., {Inutsuka}, S.-I., and {Matsumoto}, T. (2011).
\newblock {Effect of Magnetic Braking on Circumstellar Disk Formation in a
  Strongly Magnetized Cloud}.
\newblock \emph{\pasj} 63, 555--.
\newblock \doi{10.1093/pasj/63.3.555}
\bibAnnoteFile{MachidaInutsukaMatsumoto2011}

\bibitem[{{Machida} et~al.(2014){Machida}, {Inutsuka}, and
  {Matsumoto}}]{MachidaInutsukaMatsumoto2014}
{Machida}, M.~N., {Inutsuka}, S.-i., and {Matsumoto}, T. (2014).
\newblock {Conditions for circumstellar disc formation: effects of initial
  cloud configuration and sink treatment}.
\newblock \emph{\mnras} 438, 2278--2306.
\newblock \doi{10.1093/mnras/stt2343}
\bibAnnoteFile{MachidaInutsukaMatsumoto2014}

\bibitem[{{Machida} et~al.(2006){Machida}, {Matsumoto}, {Hanawa}, and
  {Tomisaka}}]{Machida+2006}
{Machida}, M.~N., {Matsumoto}, T., {Hanawa}, T., and {Tomisaka}, K. (2006).
\newblock {Evolution of Rotating Molecular Cloud Core with Oblique Magnetic
  Field}.
\newblock \emph{\apj} 645, 1227--1245.
\newblock \doi{10.1086/504423}
\bibAnnoteFile{Machida+2006}

\bibitem[{{Machida} et~al.(2004){Machida}, {Tomisaka}, and
  {Matsumoto}}]{MachidaTomisakaMatsumoto2004}
{Machida}, M.~N., {Tomisaka}, K., and {Matsumoto}, T. (2004).
\newblock {First MHD simulation of collapse and fragmentation of magnetized
  molecular cloud cores}.
\newblock \emph{\mnras} 348, L1--L5.
\newblock \doi{10.1111/j.1365-2966.2004.07402.x}
\bibAnnoteFile{MachidaTomisakaMatsumoto2004}

\bibitem[{{Marchand} et~al.(2016){Marchand}, {Masson}, {Chabrier},
  {Hennebelle}, {Commer{\c c}on}, and {Vaytet}}]{Marchand+2016}
{Marchand}, P., {Masson}, J., {Chabrier}, G., {Hennebelle}, P., {Commer{\c
  c}on}, B., and {Vaytet}, N. (2016).
\newblock {Chemical solver to compute molecule and grain abundances and
  non-ideal MHD resistivities in prestellar core-collapse calculations}.
\newblock \emph{\aap} 592, A18.
\newblock \doi{10.1051/0004-6361/201526780}
\bibAnnoteFile{Marchand+2016}

\bibitem[{{Masson} et~al.(2016){Masson}, {Chabrier}, {Hennebelle}, {Vaytet},
  and {Commer{\c c}on}}]{Masson+2016}
{Masson}, J., {Chabrier}, G., {Hennebelle}, P., {Vaytet}, N., and {Commer{\c
  c}on}, B. (2016).
\newblock {Ambipolar diffusion in low-mass star formation. I. General
  comparison with the ideal magnetohydrodynamic case}.
\newblock \emph{\aap} 587, A32.
\newblock \doi{10.1051/0004-6361/201526371}
\bibAnnoteFile{Masson+2016}

\bibitem[{{Mathis} et~al.(1977){Mathis}, {Rumpl}, and
  {Nordsieck}}]{MathisRumplNordsieck1977}
{Mathis}, J.~S., {Rumpl}, W., and {Nordsieck}, K.~H. (1977).
\newblock {The size distribution of interstellar grains}.
\newblock \emph{\apj} 217, 425--433.
\newblock \doi{10.1086/155591}
\bibAnnoteFile{MathisRumplNordsieck1977}

\bibitem[{{Matsumoto} and {Hanawa}(2011)}]{MatsumotoHanawa2011}
{Matsumoto}, T. and {Hanawa}, T. (2011).
\newblock {Protostellar Collapse of Magneto-turbulent Cloud Cores: Shape During
  Collapse and Outflow Formation}.
\newblock \emph{\apj} 728, 47.
\newblock \doi{10.1088/0004-637X/728/1/47}
\bibAnnoteFile{MatsumotoHanawa2011}

\bibitem[{{Matsumoto} et~al.(2017){Matsumoto}, {Machida}, and
  {Inutsuka}}]{MatsumotoMachidaInutsuka2017}
{Matsumoto}, T., {Machida}, M.~N., and {Inutsuka}, S.-i. (2017).
\newblock {Circumstellar Disks and Outflows in Turbulent Molecular Cloud Cores:
  Possible Formation Mechanism for Misaligned Systems}.
\newblock \emph{\apj} 839, 69.
\newblock \doi{10.3847/1538-4357/aa6a1c}
\bibAnnoteFile{MatsumotoMachidaInutsuka2017}

\bibitem[{{Matsumoto} et~al.(2006){Matsumoto}, {Nakazato}, and
  {Tomisaka}}]{MatsumotoNakazatoTomisaka2006}
{Matsumoto}, T., {Nakazato}, T., and {Tomisaka}, K. (2006).
\newblock {Alignment of Outflows with Magnetic Fields in Cloud Cores}.
\newblock \emph{\apjl} 637, L105--L108.
\newblock \doi{10.1086/500646}
\bibAnnoteFile{MatsumotoNakazatoTomisaka2006}

\bibitem[{{Matsumoto} and {Tomisaka}(2004)}]{MatsumotoTomisaka2004}
{Matsumoto}, T. and {Tomisaka}, K. (2004).
\newblock {Directions of Outflows, Disks, Magnetic Fields, and Rotation of
  Young Stellar Objects in Collapsing Molecular Cloud Cores}.
\newblock \emph{\apj} 616, 266--282.
\newblock \doi{10.1086/424897}
\bibAnnoteFile{MatsumotoTomisaka2004}

\bibitem[{{Maury} et~al.(2010){Maury}, {Andr{\'e}}, {Hennebelle}, {Motte},
  {Stamatellos}, {Bate} et~al.}]{Maury+2010}
{Maury}, A.~J., {Andr{\'e}}, P., {Hennebelle}, P., {Motte}, F., {Stamatellos},
  D., {Bate}, M., et~al. (2010).
\newblock {Toward understanding the formation of multiple systems. A pilot
  IRAM-PdBI survey of Class 0 objects}.
\newblock \emph{\aap} 512, A40.
\newblock \doi{10.1051/0004-6361/200913492}
\bibAnnoteFile{Maury+2010}

\bibitem[{{Maury} et~al.(2018){Maury}, {Girart}, {Zhang}, {Hennebelle}, {Keto},
  {Rao} et~al.}]{Maury+2018}
{Maury}, A.~J., {Girart}, J.~M., {Zhang}, Q., {Hennebelle}, P., {Keto}, E.,
  {Rao}, R., et~al. (2018).
\newblock {Magnetically regulated collapse in the B335 protostar? I. ALMA
  observations of the polarized dust emission}.
\newblock \emph{\mnras} 477, 2760--2765.
\newblock \doi{10.1093/mnras/sty574}
\bibAnnoteFile{Maury+2018}

\bibitem[{{McElroy} et~al.(2013){McElroy}, {Walsh}, {Markwick}, {Cordiner},
  {Smith}, and {Millar}}]{Mcelroy+2013}
{McElroy}, D., {Walsh}, C., {Markwick}, A.~J., {Cordiner}, M.~A., {Smith}, K.,
  and {Millar}, T.~J. (2013).
\newblock {The UMIST database for astrochemistry 2012}.
\newblock \emph{\aap} 550, A36.
\newblock \doi{10.1051/0004-6361/201220465}
\bibAnnoteFile{Mcelroy+2013}

\bibitem[{{McKee} and {Ostriker}(2007)}]{MckeeOstriker2007}
{McKee}, C.~F. and {Ostriker}, E.~C. (2007).
\newblock {Theory of Star Formation}.
\newblock \emph{\araa} 45, 565--687.
\newblock \doi{10.1146/annurev.astro.45.051806.110602}
\bibAnnoteFile{MckeeOstriker2007}

\bibitem[{{Mellon} and {Li}(2008)}]{MellonLi2008}
{Mellon}, R.~R. and {Li}, Z.-Y. (2008).
\newblock {Magnetic Braking and Protostellar Disk Formation: The Ideal MHD
  Limit}.
\newblock \emph{\apj} 681, 1356-1376.
\newblock \doi{10.1086/587542}
\bibAnnoteFile{MellonLi2008}

\bibitem[{{Mellon} and {Li}(2009)}]{MellonLi2009}
{Mellon}, R.~R. and {Li}, Z.-Y. (2009).
\newblock {Magnetic Braking and Protostellar Disk Formation: Ambipolar
  Diffusion}.
\newblock \emph{\apj} 698, 922--927.
\newblock \doi{10.1088/0004-637X/698/1/922}
\bibAnnoteFile{MellonLi2009}

\bibitem[{{Mestel} and {Spitzer}(1956)}]{MestelSpitzer1956}
{Mestel}, L. and {Spitzer}, L., Jr. (1956).
\newblock {Star formation in magnetic dust clouds}.
\newblock \emph{\mnras} 116, 503.
\newblock \doi{10.1093/mnras/116.5.503}
\bibAnnoteFile{MestelSpitzer1956}

\bibitem[{{Mouschovias}(1976)}]{Mouschovias1976b}
{Mouschovias}, T.~C. (1976).
\newblock {Nonhomologous contraction and equilibria of self-gravitating,
  magnetic interstellar clouds embedded in an intercloud medium: Star
  formation. II - Results}.
\newblock \emph{\apj} 207, 141--158.
\newblock \doi{10.1086/154478}
\bibAnnoteFile{Mouschovias1976b}

\bibitem[{{Mouschovias}(1977)}]{Mouschovias1977}
{Mouschovias}, T.~C. (1977).
\newblock {A connection between the rate of rotation of interstellar clouds,
  magnetic fields, ambipolar diffusion, and the periods of binary stars}.
\newblock \emph{\apj} 211, 147--151.
\newblock \doi{10.1086/154912}
\bibAnnoteFile{Mouschovias1977}

\bibitem[{{Mouschovias}(1979)}]{Mouschovias1979b}
{Mouschovias}, T.~C. (1979).
\newblock {Ambipolar diffusion in interstellar clouds - A new solution}.
\newblock \emph{\apj} 228, 475--481.
\newblock \doi{10.1086/156868}
\bibAnnoteFile{Mouschovias1979b}

\bibitem[{{Mouschovias}(1983)}]{Mouschovias1983}
{Mouschovias}, T.~C. (1983).
\newblock {Magnetic braking and angular momenta of protostars}.
\newblock In \emph{Solar and Stellar Magnetic Fields: Origins and Coronal
  Effects}, ed. J.~O. {Stenflo}. vol. 102 of \emph{IAU Symposium}, 479--484
\bibAnnoteFile{Mouschovias1983}

\bibitem[{{Mouschovias} and {Paleologou}(1979)}]{MouschoviasPaleologou1979}
{Mouschovias}, T.~C. and {Paleologou}, E.~V. (1979).
\newblock {The angular momentum problem and magnetic braking - an exact
  time-dependent solution}.
\newblock \emph{\apj} 230, 204--222.
\newblock \doi{10.1086/157077}
\bibAnnoteFile{MouschoviasPaleologou1979}

\bibitem[{{Mouschovias} and {Paleologou}(1980)}]{MouschoviasPaleologou1980}
{Mouschovias}, T.~C. and {Paleologou}, E.~V. (1980).
\newblock {Magnetic braking of an aligned rotator during star formation - an
  exact, time-dependent solution}.
\newblock \emph{\apj} 237, 877--899.
\newblock \doi{10.1086/157936}
\bibAnnoteFile{MouschoviasPaleologou1980}

\bibitem[{{Mouschovias} and {Spitzer}(1976)}]{MouschoviasSpitzer1976}
{Mouschovias}, T.~C. and {Spitzer}, L., Jr. (1976).
\newblock {Note on the collapse of magnetic interstellar clouds}.
\newblock \emph{\apj} 210, 326.
\newblock \doi{10.1086/154835}
\bibAnnoteFile{MouschoviasSpitzer1976}

\bibitem[{{Murillo} et~al.(2013){Murillo}, {Lai}, {Bruderer}, {Harsono}, and
  {van Dishoeck}}]{Murillo+2013}
{Murillo}, N.~M., {Lai}, S.-P., {Bruderer}, S., {Harsono}, D., and {van
  Dishoeck}, E.~F. (2013).
\newblock {A Keplerian disk around a Class 0 source: ALMA observations of
  VLA1623A}.
\newblock \emph{\aap} 560, A103.
\newblock \doi{10.1051/0004-6361/201322537}
\bibAnnoteFile{Murillo+2013}

\bibitem[{{Myers} et~al.(2013){Myers}, {McKee}, {Cunningham}, {Klein}, and
  {Krumholz}}]{Myers+2013}
{Myers}, A.~T., {McKee}, C.~F., {Cunningham}, A.~J., {Klein}, R.~I., and
  {Krumholz}, M.~R. (2013).
\newblock {The Fragmentation of Magnetized, Massive Star-forming Cores with
  Radiative Feedback}.
\newblock \emph{\apj} 766, 97.
\newblock \doi{10.1088/0004-637X/766/2/97}
\bibAnnoteFile{Myers+2013}

\bibitem[{{Nakano} and {Nakamura}(1978)}]{NakanoNakamura1978}
{Nakano}, T. and {Nakamura}, T. (1978).
\newblock {Gravitational Instability of Magnetized Gaseous Disks 6}.
\newblock \emph{\pasj} 30, 671--680
\bibAnnoteFile{NakanoNakamura1978}

\bibitem[{{Nakano} et~al.(2002){Nakano}, {Nishi}, and
  {Umebayashi}}]{NakanoNishiUmebayashi2002}
{Nakano}, T., {Nishi}, R., and {Umebayashi}, T. (2002).
\newblock {Mechanism of Magnetic Flux Loss in Molecular Clouds}.
\newblock \emph{\apj} 573, 199--214.
\newblock \doi{10.1086/340587}
\bibAnnoteFile{NakanoNishiUmebayashi2002}

\bibitem[{{Nakano} and {Umebayashi}(1986)}]{NakanoUmebayashi1986}
{Nakano}, T. and {Umebayashi}, T. (1986).
\newblock {Dissipation of magnetic fields in very dense interstellar clouds. I
  - Formulation and conditions for efficient dissipation}.
\newblock \emph{\mnras} 218, 663--684.
\newblock \doi{10.1093/mnras/218.4.663}
\bibAnnoteFile{NakanoUmebayashi1986}

\bibitem[{{Nishi} et~al.(1991){Nishi}, {Nakano}, and
  {Umebayashi}}]{NishiNakanoUmebayashi1991}
{Nishi}, R., {Nakano}, T., and {Umebayashi}, T. (1991).
\newblock {Magnetic flux loss from interstellar clouds with various grain-size
  distributions}.
\newblock \emph{\apj} 368, 181--194.
\newblock \doi{10.1086/169682}
\bibAnnoteFile{NishiNakanoUmebayashi1991}

\bibitem[{{Oka} et~al.(2005){Oka}, {Geballe}, {Goto}, {Usuda}, and
  {McCall}}]{Oka+2005}
{Oka}, T., {Geballe}, T.~R., {Goto}, M., {Usuda}, T., and {McCall}, B.~J.
  (2005).
\newblock {Hot and Diffuse Clouds near the Galactic Center Probed by Metastable
  H$^{+}$$_{3}$1,}.
\newblock \emph{\apj} 632, 882--893.
\newblock \doi{10.1086/432679}
\bibAnnoteFile{Oka+2005}

\bibitem[{{Okuzumi}(2009)}]{Okuzumi2009}
{Okuzumi}, S. (2009).
\newblock {Electric Charging of Dust Aggregates and its Effect on Dust
  Coagulation in Protoplanetary Disks}.
\newblock \emph{\apj} 698, 1122--1135.
\newblock \doi{10.1088/0004-637X/698/2/1122}
\bibAnnoteFile{Okuzumi2009}

\bibitem[{{O'Sullivan} and {Downes}(2006)}]{OsullivanDownes2006}
{O'Sullivan}, S. and {Downes}, T.~P. (2006).
\newblock {An explicit scheme for multifluid magnetohydrodynamics}.
\newblock \emph{\mnras} 366, 1329--1336.
\newblock \doi{10.1111/j.1365-2966.2005.09898.x}
\bibAnnoteFile{OsullivanDownes2006}

\bibitem[{{Padoan} and {Nordlund}(2002)}]{PadoanNordlund2002}
{Padoan}, P. and {Nordlund}, {\AA}. (2002).
\newblock {The Stellar Initial Mass Function from Turbulent Fragmentation}.
\newblock \emph{\apj} 576, 870--879.
\newblock \doi{10.1086/341790}
\bibAnnoteFile{PadoanNordlund2002}

\bibitem[{{Padovani} et~al.(2009){Padovani}, {Galli}, and
  {Glassgold}}]{PadovaniGalliGlassgold2009}
{Padovani}, M., {Galli}, D., and {Glassgold}, A.~E. (2009).
\newblock {Cosmic-ray ionization of molecular clouds}.
\newblock \emph{\aap} 501, 619--631.
\newblock \doi{10.1051/0004-6361/200911794}
\bibAnnoteFile{PadovaniGalliGlassgold2009}

\bibitem[{{Padovani} et~al.(2014){Padovani}, {Galli}, {Hennebelle}, {Commer{\c
  c}on}, and {Joos}}]{Padovani+2014}
{Padovani}, M., {Galli}, D., {Hennebelle}, P., {Commer{\c c}on}, B., and
  {Joos}, M. (2014).
\newblock {The role of cosmic rays on magnetic field diffusion and the
  formation of protostellar discs}.
\newblock \emph{\aap} 571, A33.
\newblock \doi{10.1051/0004-6361/201424035}
\bibAnnoteFile{Padovani+2014}

\bibitem[{{Planck Collaboration} et~al.(2015){Planck Collaboration}, {Ade},
  {Aghanim}, {Alina}, {Alves}, {Armitage-Caplan} et~al.}]{Planck2015}
{Planck Collaboration}, {Ade}, P.~A.~R., {Aghanim}, N., {Alina}, D., {Alves},
  M.~I.~R., {Armitage-Caplan}, C., et~al. (2015).
\newblock {Planck intermediate results. XIX. An overview of the polarized
  thermal emission from Galactic dust}.
\newblock \emph{\aap} 576, A104.
\newblock \doi{10.1051/0004-6361/201424082}
\bibAnnoteFile{Planck2015}

\bibitem[{{Price} and {Bate}(2007)}]{PriceBate2007}
{Price}, D.~J. and {Bate}, M.~R. (2007).
\newblock {The impact of magnetic fields on single and binary star formation}.
\newblock \emph{\mnras} 377, 77--90.
\newblock \doi{10.1111/j.1365-2966.2007.11621.x}
\bibAnnoteFile{PriceBate2007}

\bibitem[{{Price} et~al.(2018){Price}, {Wurster}, {Tricco}, {Nixon}, {Toupin},
  {Pettitt} et~al.}]{Phantom2018}
{Price}, D.~J., {Wurster}, J., {Tricco}, T.~S., {Nixon}, C., {Toupin}, S.,
  {Pettitt}, A., et~al. (2018).
\newblock {Phantom: A Smoothed Particle Hydrodynamics and Magnetohydrodynamics
  Code for Astrophysics}.
\newblock \emph{\pasa} 35, e031.
\newblock \doi{10.1017/pasa.2018.25}
\bibAnnoteFile{Phantom2018}

\bibitem[{{Rao} et~al.(2014){Rao}, {Girart}, {Lai}, and {Marrone}}]{Rao+2014}
{Rao}, R., {Girart}, J.~M., {Lai}, S.-P., and {Marrone}, D.~P. (2014).
\newblock {Detection of a Magnetized Disk around a Very Young Protostar}.
\newblock \emph{\apjl} 780, L6.
\newblock \doi{10.1088/2041-8205/780/1/L6}
\bibAnnoteFile{Rao+2014}

\bibitem[{{Rodgers-Lee} et~al.(2016){Rodgers-Lee}, {Ray}, and
  {Downes}}]{RodgersleeRayDownes2016}
{Rodgers-Lee}, D., {Ray}, T.~P., and {Downes}, T.~P. (2016).
\newblock {Global multifluid simulations of the magnetorotational instability
  in radially stratified protoplanetary discs}.
\newblock \emph{\mnras} 463, 134--145.
\newblock \doi{10.1093/mnras/stw1980}
\bibAnnoteFile{RodgersleeRayDownes2016}

\bibitem[{{Sadavoy} et~al.(2018){Sadavoy}, {Myers}, {Stephens}, {Tobin},
  {Commer{\c c}on}, {Henning} et~al.}]{Sadavoy+2018}
{Sadavoy}, S.~I., {Myers}, P.~C., {Stephens}, I.~W., {Tobin}, J., {Commer{\c
  c}on}, B., {Henning}, T., et~al. (2018).
\newblock {Dust Polarization toward Embedded Protostars in Ophiuchus with ALMA.
  I. VLA 1623}.
\newblock \emph{\apj} 859, 165.
\newblock \doi{10.3847/1538-4357/aac21a}
\bibAnnoteFile{Sadavoy+2018}

\bibitem[{{Saigo} and {Tomisaka}(2006)}]{SaigoTomisaka2006}
{Saigo}, K. and {Tomisaka}, K. (2006).
\newblock {Evolution of First Cores in Rotating Molecular Cores}.
\newblock \emph{\apj} 645, 381--394.
\newblock \doi{10.1086/504028}
\bibAnnoteFile{SaigoTomisaka2006}

\bibitem[{{Saigo} et~al.(2008){Saigo}, {Tomisaka}, and
  {Matsumoto}}]{SaigoTomisakaMatsumoto2008}
{Saigo}, K., {Tomisaka}, K., and {Matsumoto}, T. (2008).
\newblock {Evolution of First Cores and Formation of Stellar Cores in Rotating
  Molecular Cloud Cores}.
\newblock \emph{\apj} 674, 997--1014.
\newblock \doi{10.1086/523888}
\bibAnnoteFile{SaigoTomisakaMatsumoto2008}

\bibitem[{{Sano} and {Stone}(2002{\natexlab{a}})}]{SanoStone2002a}
{Sano}, T. and {Stone}, J.~M. (2002{\natexlab{a}}).
\newblock {The Effect of the Hall Term on the Nonlinear Evolution of the
  Magnetorotational Instability. I. Local Axisymmetric Simulations}.
\newblock \emph{\apj} 570, 314--328.
\newblock \doi{10.1086/339504}
\bibAnnoteFile{SanoStone2002a}

\bibitem[{{Sano} and {Stone}(2002{\natexlab{b}})}]{SanoStone2002b}
{Sano}, T. and {Stone}, J.~M. (2002{\natexlab{b}}).
\newblock {The Effect of the Hall Term on the Nonlinear Evolution of the
  Magnetorotational Instability. II. Saturation Level and Critical Magnetic
  Reynolds Number}.
\newblock \emph{\apj} 577, 534--553.
\newblock \doi{10.1086/342172}
\bibAnnoteFile{SanoStone2002b}

\bibitem[{{Santos-Lima} et~al.(2012){Santos-Lima}, {de Gouveia Dal Pino}, and
  {Lazarian}}]{Santoslima+2012}
{Santos-Lima}, R., {de Gouveia Dal Pino}, E.~M., and {Lazarian}, A. (2012).
\newblock {The Role of Turbulent Magnetic Reconnection in the Formation of
  Rotationally Supported Protostellar Disks}.
\newblock \emph{\apj} 747, 21.
\newblock \doi{10.1088/0004-637X/747/1/21}
\bibAnnoteFile{Santoslima+2012}

\bibitem[{{Santos-Lima} et~al.(2013){Santos-Lima}, {de Gouveia Dal Pino}, and
  {Lazarian}}]{Santoslima+2013}
{Santos-Lima}, R., {de Gouveia Dal Pino}, E.~M., and {Lazarian}, A. (2013).
\newblock {Disc formation in turbulent cloud cores: is magnetic flux loss
  necessary to stop the magnetic braking catastrophe or not?}
\newblock \emph{\mnras} 429, 3371--3378.
\newblock \doi{10.1093/mnras/sts597}
\bibAnnoteFile{Santoslima+2013}

\bibitem[{{Segura-Cox} et~al.(2016){Segura-Cox}, {Harris}, {Tobin}, {Looney},
  {Li}, {Chandler} et~al.}]{Seguracox+2016}
{Segura-Cox}, D.~M., {Harris}, R.~J., {Tobin}, J.~J., {Looney}, L.~W., {Li},
  Z.-Y., {Chandler}, C., et~al. (2016).
\newblock {The VLA Nascent Disk and Multiplicity Survey: First Look at Resolved
  Candidate Disks around Class 0 and I Protostars in the Perseus Molecular
  Cloud}.
\newblock \emph{\apjl} 817, L14.
\newblock \doi{10.3847/2041-8205/817/2/L14}
\bibAnnoteFile{Seguracox+2016}

\bibitem[{{Segura-Cox} et~al.(2015){Segura-Cox}, {Looney}, {Stephens},
  {Fern{\'a}ndez-L{\'o}pez}, {Kwon}, {Tobin} et~al.}]{Seguracox+2015}
{Segura-Cox}, D.~M., {Looney}, L.~W., {Stephens}, I.~W.,
  {Fern{\'a}ndez-L{\'o}pez}, M., {Kwon}, W., {Tobin}, J.~J., et~al. (2015).
\newblock {The Magnetic Field in the Class 0 Protostellar Disk of L1527}.
\newblock \emph{\apjl} 798, L2.
\newblock \doi{10.1088/2041-8205/798/1/L2}
\bibAnnoteFile{Seguracox+2015}

\bibitem[{{Seifried} et~al.(2012){Seifried}, {Banerjee}, {Pudritz}, and
  {Klessen}}]{Seifried+2012}
{Seifried}, D., {Banerjee}, R., {Pudritz}, R.~E., and {Klessen}, R.~S. (2012).
\newblock {Disc formation in turbulent massive cores: circumventing the
  magnetic braking catastrophe}.
\newblock \emph{\mnras} 423, L40--L44.
\newblock \doi{10.1111/j.1745-3933.2012.01253.x}
\bibAnnoteFile{Seifried+2012}

\bibitem[{{Seifried} et~al.(2013){Seifried}, {Banerjee}, {Pudritz}, and
  {Klessen}}]{Seifried+2013}
{Seifried}, D., {Banerjee}, R., {Pudritz}, R.~E., and {Klessen}, R.~S. (2013).
\newblock {Turbulence-induced disc formation in strongly magnetized cloud
  cores}.
\newblock \emph{\mnras} 432, 3320--3331.
\newblock \doi{10.1093/mnras/stt682}
\bibAnnoteFile{Seifried+2013}

\bibitem[{{Shu} et~al.(2006){Shu}, {Galli}, {Lizano}, and {Cai}}]{Shu+2006}
{Shu}, F.~H., {Galli}, D., {Lizano}, S., and {Cai}, M. (2006).
\newblock {Gravitational Collapse of Magnetized Clouds. II. The Role of Ohmic
  Dissipation}.
\newblock \emph{\apj} 647, 382--389.
\newblock \doi{10.1086/505258}
\bibAnnoteFile{Shu+2006}

\bibitem[{{Spitzer}(1968)}]{Spitzer1968}
{Spitzer}, L., Jr. (1968).
\newblock \emph{{Dynamics of Interstellar Matter and the Formation of Stars}}
  (the University of Chicago Press).
\newblock 1
\bibAnnoteFile{Spitzer1968}

\bibitem[{{Spitzer} and {Tomasko}(1968)}]{SpitzerTomasko1968}
{Spitzer}, L., Jr. and {Tomasko}, M.~G. (1968).
\newblock {Heating of H i Regions by Energetic Particles}.
\newblock \emph{\apj} 152, 971.
\newblock \doi{10.1086/149610}
\bibAnnoteFile{SpitzerTomasko1968}

\bibitem[{{Stephens} et~al.(2014){Stephens}, {Looney}, {Kwon},
  {Fern{\'a}ndez-L{\'o}pez}, {Hughes}, {Mundy} et~al.}]{Stephens+2014}
{Stephens}, I.~W., {Looney}, L.~W., {Kwon}, W., {Fern{\'a}ndez-L{\'o}pez}, M.,
  {Hughes}, A.~M., {Mundy}, L.~G., et~al. (2014).
\newblock {Spatially resolved magnetic field structure in the disk of a T Tauri
  star}.
\newblock \emph{\nat} 514, 597--599.
\newblock \doi{10.1038/nature13850}
\bibAnnoteFile{Stephens+2014}

\bibitem[{{Stephens} et~al.(2017){Stephens}, {Yang}, {Li}, {Looney}, {Kataoka},
  {Kwon} et~al.}]{Stephens+2017}
{Stephens}, I.~W., {Yang}, H., {Li}, Z.-Y., {Looney}, L.~W., {Kataoka}, A.,
  {Kwon}, W., et~al. (2017).
\newblock {ALMA Reveals Transition of Polarization Pattern with Wavelength in
  HL Tau's Disk}.
\newblock \emph{\apj} 851, 55.
\newblock \doi{10.3847/1538-4357/aa998b}
\bibAnnoteFile{Stephens+2017}

\bibitem[{{Tassis} and
  {Mouschovias}(2005{\natexlab{a}})}]{TassisMouschovias2005a}
{Tassis}, K. and {Mouschovias}, T.~C. (2005{\natexlab{a}}).
\newblock {Magnetically Controlled Spasmodic Accretion during Star Formation.
  I. Formulation of the Problem and Method of Solution}.
\newblock \emph{\apj} 618, 769--782.
\newblock \doi{10.1086/424479}
\bibAnnoteFile{TassisMouschovias2005a}

\bibitem[{{Tassis} and
  {Mouschovias}(2005{\natexlab{b}})}]{TassisMouschovias2005b}
{Tassis}, K. and {Mouschovias}, T.~C. (2005{\natexlab{b}}).
\newblock {Magnetically Controlled Spasmodic Accretion during Star Formation.
  II. Results}.
\newblock \emph{\apj} 618, 783--794.
\newblock \doi{10.1086/424480}
\bibAnnoteFile{TassisMouschovias2005b}

\bibitem[{{Tassis} and
  {Mouschovias}(2007{\natexlab{a}})}]{TassisMouschovias2007a}
{Tassis}, K. and {Mouschovias}, T.~C. (2007{\natexlab{a}}).
\newblock {Protostar Formation in Magnetic Molecular Clouds beyond Ion
  Detachment. I. Formulation of the Problem and Method of Solution}.
\newblock \emph{\apj} 660, 370--387.
\newblock \doi{10.1086/512760}
\bibAnnoteFile{TassisMouschovias2007a}

\bibitem[{{Tassis} and
  {Mouschovias}(2007{\natexlab{b}})}]{TassisMouschovias2007b}
{Tassis}, K. and {Mouschovias}, T.~C. (2007{\natexlab{b}}).
\newblock {Protostar Formation in Magnetic Molecular Clouds beyond Ion
  Detachment. II. Typical Axisymmetric Solution}.
\newblock \emph{\apj} 660, 388--401.
\newblock \doi{10.1086/512761}
\bibAnnoteFile{TassisMouschovias2007b}

\bibitem[{{Tassis} and
  {Mouschovias}(2007{\natexlab{c}})}]{TassisMouschovias2007c}
{Tassis}, K. and {Mouschovias}, T.~C. (2007{\natexlab{c}}).
\newblock {Protostar Formation in Magnetic Molecular Clouds beyond Ion
  Detachment. III. A Parameter Study}.
\newblock \emph{\apj} 660, 402--417.
\newblock \doi{10.1086/512762}
\bibAnnoteFile{TassisMouschovias2007c}

\bibitem[{{Tobin} et~al.(2012){Tobin}, {Hartmann}, {Chiang}, {Wilner},
  {Looney}, {Loinard} et~al.}]{Tobin+2012}
{Tobin}, J.~J., {Hartmann}, L., {Chiang}, H.-F., {Wilner}, D.~J., {Looney},
  L.~W., {Loinard}, L., et~al. (2012).
\newblock {A \~{}0.2-solar-mass protostar with a Keplerian disk in the very
  young L1527 IRS system}.
\newblock \emph{\nat} 492, 83--85.
\newblock \doi{10.1038/nature11610}
\bibAnnoteFile{Tobin+2012}

\bibitem[{{Tomida}(2014)}]{Tomida2014}
{Tomida}, K. (2014).
\newblock {Radiation Magnetohydrodynamic Simulations of Protostellar Collapse:
  Low-metallicity Environments}.
\newblock \emph{\apj} 786, 98.
\newblock \doi{10.1088/0004-637X/786/2/98}
\bibAnnoteFile{Tomida2014}

\bibitem[{{Tomida} et~al.(2017){Tomida}, {Machida}, {Hosokawa}, {Sakurai}, and
  {Lin}}]{Tomida+2017}
{Tomida}, K., {Machida}, M.~N., {Hosokawa}, T., {Sakurai}, Y., and {Lin}, C.~H.
  (2017).
\newblock {Grand-design Spiral Arms in a Young Forming Circumstellar Disk}.
\newblock \emph{\apjl} 835, L11.
\newblock \doi{10.3847/2041-8213/835/1/L11}
\bibAnnoteFile{Tomida+2017}

\bibitem[{{Tomida} et~al.(2015){Tomida}, {Okuzumi}, and
  {Machida}}]{TomidaOkuzumiMachida2015}
{Tomida}, K., {Okuzumi}, S., and {Machida}, M.~N. (2015).
\newblock {Radiation Magnetohydrodynamic Simulations of Protostellar Collapse:
  Nonideal Magnetohydrodynamic Effects and Early Formation of Circumstellar
  Disks}.
\newblock \emph{\apj} 801, 117.
\newblock \doi{10.1088/0004-637X/801/2/117}
\bibAnnoteFile{TomidaOkuzumiMachida2015}

\bibitem[{{Tomida} et~al.(2013){Tomida}, {Tomisaka}, {Matsumoto}, {Hori},
  {Okuzumi}, {Machida} et~al.}]{Tomida+2013}
{Tomida}, K., {Tomisaka}, K., {Matsumoto}, T., {Hori}, Y., {Okuzumi}, S.,
  {Machida}, M.~N., et~al. (2013).
\newblock {Radiation Magnetohydrodynamic Simulations of Protostellar Collapse:
  Protostellar Core Formation}.
\newblock \emph{\apj} 763, 6.
\newblock \doi{10.1088/0004-637X/763/1/6}
\bibAnnoteFile{Tomida+2013}

\bibitem[{{Tomida} et~al.(2010){Tomida}, {Tomisaka}, {Matsumoto}, {Ohsuga},
  {Machida}, and {Saigo}}]{Tomida+2010rmhd}
{Tomida}, K., {Tomisaka}, K., {Matsumoto}, T., {Ohsuga}, K., {Machida}, M.~N.,
  and {Saigo}, K. (2010).
\newblock {Radiation Magnetohydrodynamics Simulation of Proto-stellar Collapse:
  Two-component Molecular Outflow}.
\newblock \emph{\apjl} 714, L58--L63.
\newblock \doi{10.1088/2041-8205/714/1/L58}
\bibAnnoteFile{Tomida+2010rmhd}

\bibitem[{{Tomisaka}(2000)}]{Tomisaka2000}
{Tomisaka}, K. (2000).
\newblock {The Evolution of the Angular Momentum Distribution during Star
  Formation}.
\newblock \emph{\apjl} 528, L41--L44.
\newblock \doi{10.1086/312417}
\bibAnnoteFile{Tomisaka2000}

\bibitem[{{Toomre}(1964)}]{Toomre1964}
{Toomre}, A. (1964).
\newblock {On the gravitational stability of a disk of stars}.
\newblock \emph{\apj} 139, 1217--1238.
\newblock \doi{10.1086/147861}
\bibAnnoteFile{Toomre1964}

\bibitem[{{Troland} and {Crutcher}(2008)}]{TrolandCrutcher2008}
{Troland}, T.~H. and {Crutcher}, R.~M. (2008).
\newblock {Magnetic Fields in Dark Cloud Cores: Arecibo OH Zeeman
  Observations}.
\newblock \emph{\apj} 680, 457-465.
\newblock \doi{10.1086/587546}
\bibAnnoteFile{TrolandCrutcher2008}

\bibitem[{{Tsukamoto} et~al.(2015{\natexlab{a}}){Tsukamoto}, {Iwasaki},
  {Okuzumi}, {Machida}, and {Inutsuka}}]{Tsukamoto+2015hall}
{Tsukamoto}, Y., {Iwasaki}, K., {Okuzumi}, S., {Machida}, M.~N., and
  {Inutsuka}, S. (2015{\natexlab{a}}).
\newblock {Bimodality of Circumstellar Disk Evolution Induced by the Hall
  Current}.
\newblock \emph{\apjl} 810, L26.
\newblock \doi{10.1088/2041-8205/810/2/L26}
\bibAnnoteFile{Tsukamoto+2015hall}

\bibitem[{{Tsukamoto} et~al.(2015{\natexlab{b}}){Tsukamoto}, {Iwasaki},
  {Okuzumi}, {Machida}, and {Inutsuka}}]{Tsukamoto+2015oa}
{Tsukamoto}, Y., {Iwasaki}, K., {Okuzumi}, S., {Machida}, M.~N., and
  {Inutsuka}, S. (2015{\natexlab{b}}).
\newblock {Effects of Ohmic and ambipolar diffusion on formation and evolution
  of first cores, protostars, and circumstellar discs}.
\newblock \emph{\mnras} 452, 278--288.
\newblock \doi{10.1093/mnras/stv1290}
\bibAnnoteFile{Tsukamoto+2015oa}

\bibitem[{{Tsukamoto} et~al.(2017){Tsukamoto}, {Okuzumi}, {Iwasaki}, {Machida},
  and {Inutsuka}}]{Tsukamoto+2017}
{Tsukamoto}, Y., {Okuzumi}, S., {Iwasaki}, K., {Machida}, M.~N., and
  {Inutsuka}, S.-i. (2017).
\newblock {The impact of the Hall effect during cloud core collapse:
  Implications for circumstellar disk evolution}.
\newblock \emph{\pasj} 69, 95.
\newblock \doi{10.1093/pasj/psx113}
\bibAnnoteFile{Tsukamoto+2017}

\bibitem[{{Turner} and {Sano}(2008)}]{TurnerSano2008}
{Turner}, N.~J. and {Sano}, T. (2008).
\newblock {Dead Zone Accretion Flows in Protostellar Disks}.
\newblock \emph{\apjl} 679, L131.
\newblock \doi{10.1086/589540}
\bibAnnoteFile{TurnerSano2008}

\bibitem[{{Umebayashi} and {Nakano}(1981)}]{UmebayashiNakano1981}
{Umebayashi}, T. and {Nakano}, T. (1981).
\newblock {Fluxes of Energetic Particles and the Ionization Rate in Very Dense
  Interstellar Clouds}.
\newblock \emph{\pasj} 33, 617
\bibAnnoteFile{UmebayashiNakano1981}

\bibitem[{{Umebayashi} and {Nakano}(1990)}]{UmebayashiNakano1990}
{Umebayashi}, T. and {Nakano}, T. (1990).
\newblock {Magnetic flux loss from interstellar clouds}.
\newblock \emph{\mnras} 243, 103--113.
\newblock \doi{10.1093/mnras/243.1.103}
\bibAnnoteFile{UmebayashiNakano1990}

\bibitem[{{Umebayashi} and {Nakano}(2009)}]{UmebayashiNakano2009}
{Umebayashi}, T. and {Nakano}, T. (2009).
\newblock {Effects of Radionuclides on the Ionization State of Protoplanetary
  Disks and Dense Cloud Cores}.
\newblock \emph{\apj} 690, 69--81.
\newblock \doi{10.1088/0004-637X/690/1/69}
\bibAnnoteFile{UmebayashiNakano2009}

\bibitem[{{Vaytet} et~al.(2018){Vaytet}, {Commer{\c c}on}, {Masson},
  {Gonz{\'a}lez}, and {Chabrier}}]{Vaytet+2018}
{Vaytet}, N., {Commer{\c c}on}, B., {Masson}, J., {Gonz{\'a}lez}, M., and
  {Chabrier}, G. (2018).
\newblock {Protostellar birth with ambipolar and ohmic diffusion}.
\newblock \emph{\aap} 615, A5.
\newblock \doi{10.1051/0004-6361/201732075}
\bibAnnoteFile{Vaytet+2018}

\bibitem[{{Ward-Thompson} et~al.(2010){Ward-Thompson}, {Kirk}, {Andr{\'e}},
  {Saraceno}, {Didelon}, {K{\"o}nyves} et~al.}]{Wardthompson+2010}
{Ward-Thompson}, D., {Kirk}, J.~M., {Andr{\'e}}, P., {Saraceno}, P., {Didelon},
  P., {K{\"o}nyves}, V., et~al. (2010).
\newblock {A Herschel study of the properties of starless cores in the Polaris
  Flare dark cloud region using PACS and SPIRE}.
\newblock \emph{\aap} 518, L92.
\newblock \doi{10.1051/0004-6361/201014618}
\bibAnnoteFile{Wardthompson+2010}

\bibitem[{{Wardle}(2007)}]{Wardle2007}
{Wardle}, M. (2007).
\newblock {Magnetic fields in protoplanetary disks}.
\newblock \emph{\apss} 311, 35--45.
\newblock \doi{10.1007/s10509-007-9575-8}
\bibAnnoteFile{Wardle2007}

\bibitem[{{Wardle} and {Koenigl}(1993)}]{WardleKoenigl1993}
{Wardle}, M. and {Koenigl}, A. (1993).
\newblock {The structure of protostellar accretion disks and the origin of
  bipolar flows}.
\newblock \emph{\apj} 410, 218--238.
\newblock \doi{10.1086/172739}
\bibAnnoteFile{WardleKoenigl1993}

\bibitem[{{Wardle} and {Ng}(1999)}]{WardleNg1999}
{Wardle}, M. and {Ng}, C. (1999).
\newblock {The conductivity of dense molecular gas}.
\newblock \emph{\mnras} 303, 239--246.
\newblock \doi{10.1046/j.1365-8711.1999.02211.x}
\bibAnnoteFile{WardleNg1999}

\bibitem[{{Weiss}(1966)}]{Weiss1966}
{Weiss}, N.~O. (1966).
\newblock {The Expulsion of Magnetic Flux by Eddies}.
\newblock \emph{Proceedings of the Royal Society of London Series A} 293,
  310--328.
\newblock \doi{10.1098/rspa.1966.0173}
\bibAnnoteFile{Weiss1966}

\bibitem[{{Wurster}(2016)}]{Wurster2016}
{Wurster}, J. (2016).
\newblock {NICIL: A Stand Alone Library to Self-Consistently Calculate
  Non-Ideal Magnetohydrodynamic Coefficients in Molecular Cloud Cores}.
\newblock \emph{\pasa} 33, e041.
\newblock \doi{10.1017/pasa.2016.34}
\bibAnnoteFile{Wurster2016}

\bibitem[{{Wurster} et~al.(2018{\natexlab{a}}){Wurster}, {Bate}, and
  {Price}}]{WursterBatePrice2018sd}
{Wurster}, J., {Bate}, M.~R., and {Price}, D.~J. (2018{\natexlab{a}}).
\newblock { The collapse of a molecular cloud core to stellar densities using
  radiation non-ideal magnetohydrodynamics}.
\newblock \emph{\mnras} 475, 1859--1880.
\newblock \doi{10.1093/mnras/stx3339}
\bibAnnoteFile{WursterBatePrice2018sd}

\bibitem[{{Wurster} et~al.(2018{\natexlab{b}}){Wurster}, {Bate}, and
  {Price}}]{WursterBatePrice2018ion}
{Wurster}, J., {Bate}, M.~R., and {Price}, D.~J. (2018{\natexlab{b}}).
\newblock { The effect of extreme ionization rates during the initial collapse
  of a molecular cloud core}.
\newblock \emph{\mnras} 476, 2063--2074.
\newblock \doi{10.1093/mnras/sty392}
\bibAnnoteFile{WursterBatePrice2018ion}

\bibitem[{{Wurster} et~al.(2018{\natexlab{c}}){Wurster}, {Bate}, and
  {Price}}]{WursterBatePrice2018hd}
{Wurster}, J., {Bate}, M.~R., and {Price}, D.~J. (2018{\natexlab{c}}).
\newblock {Hall effect-driven formation of gravitationally unstable discs in
  magnetized molecular cloud cores}.
\newblock \emph{\mnras} 480, 4434--4442.
\newblock \doi{10.1093/mnras/sty2212}
\bibAnnoteFile{WursterBatePrice2018hd}

\bibitem[{{Wurster} et~al.(2018{\natexlab{d}}){Wurster}, {Bate}, and
  {Price}}]{WursterBatePrice2018ff}
{Wurster}, J., {Bate}, M.~R., and {Price}, D.~J. (2018{\natexlab{d}}).
\newblock {On the origin of magnetic fields in stars}.
\newblock \emph{\mnras} 481, 2450--2457.
\newblock \doi{10.1093/mnras/sty2438}
\bibAnnoteFile{WursterBatePrice2018ff}

\bibitem[{{Wurster} et~al.(2014){Wurster}, {Price}, and
  {Ayliffe}}]{WursterPriceAyliffe2014}
{Wurster}, J., {Price}, D.~J., and {Ayliffe}, B. (2014).
\newblock {Ambipolar diffusion in smoothed particle magnetohydrodynamics}.
\newblock \emph{\mnras} 444, 1104--1112.
\newblock \doi{10.1093/mnras/stu1524}
\bibAnnoteFile{WursterPriceAyliffe2014}

\bibitem[{{Wurster} et~al.(2016){Wurster}, {Price}, and
  {Bate}}]{WursterPriceBate2016}
{Wurster}, J., {Price}, D.~J., and {Bate}, M.~R. (2016).
\newblock {Can non-ideal magnetohydrodynamics solve the magnetic braking
  catastrophe?}
\newblock \emph{\mnras} 457, 1037--1061.
\newblock \doi{10.1093/mnras/stw013}
\bibAnnoteFile{WursterPriceBate2016}

\bibitem[{{Yang} et~al.(2016{\natexlab{a}}){Yang}, {Li}, {Looney}, and
  {Stephens}}]{Yang+2016hltau}
{Yang}, H., {Li}, Z.-Y., {Looney}, L., and {Stephens}, I. (2016{\natexlab{a}}).
\newblock {Inclination-induced polarization of scattered millimetre radiation
  from protoplanetary discs: the case of HL Tau}.
\newblock \emph{\mnras} 456, 2794--2805.
\newblock \doi{10.1093/mnras/stv2633}
\bibAnnoteFile{Yang+2016hltau}

\bibitem[{{Yang} et~al.(2016{\natexlab{b}}){Yang}, {Li}, {Looney}, {Cox},
  {Tobin}, {Stephens} et~al.}]{Yang+2016ngc}
{Yang}, H., {Li}, Z.-Y., {Looney}, L.~W., {Cox}, E.~G., {Tobin}, J.,
  {Stephens}, I.~W., et~al. (2016{\natexlab{b}}).
\newblock {Disc polarization from both emission and scattering of magnetically
  aligned grains: the case of NGC 1333 IRAS 4A1}.
\newblock \emph{\mnras} 460, 4109--4121.
\newblock \doi{10.1093/mnras/stw1253}
\bibAnnoteFile{Yang+2016ngc}

\bibitem[{{Yang} et~al.(2017){Yang}, {Li}, {Looney}, {Girart}, and
  {Stephens}}]{Yang+2017}
{Yang}, H., {Li}, Z.-Y., {Looney}, L.~W., {Girart}, J.~M., and {Stephens},
  I.~W. (2017).
\newblock {Scattering-produced (sub)millimetre polarization in inclined discs:
  optical depth effects, near-far side asymmetry and dust settling}.
\newblock \emph{\mnras} 472, 373--388.
\newblock \doi{10.1093/mnras/stx1951}
\bibAnnoteFile{Yang+2017}

\bibitem[{{Yen} et~al.(2015){Yen}, {Koch}, {Takakuwa}, {Ho}, {Ohashi}, and
  {Tang}}]{Yen+2015}
{Yen}, H.-W., {Koch}, P.~M., {Takakuwa}, S., {Ho}, P.~T.~P., {Ohashi}, N., and
  {Tang}, Y.-W. (2015).
\newblock {Observations of Infalling and Rotational Motions on a 1000 AU Scale
  around 17 Class 0 and 0/I Protostars: Hints of Disk Growth and Magnetic
  Braking?}
\newblock \emph{\apj} 799, 193.
\newblock \doi{10.1088/0004-637X/799/2/193}
\bibAnnoteFile{Yen+2015}

\bibitem[{{Yorke} et~al.(1993){Yorke}, {Bodenheimer}, and
  {Laughlin}}]{YorkeBodenheimerLaughlin1993}
{Yorke}, H.~W., {Bodenheimer}, P., and {Laughlin}, G. (1993).
\newblock {The formation of protostellar disks. I - 1 M(solar)}.
\newblock \emph{\apj} 411, 274--284.
\newblock \doi{10.1086/172827}
\bibAnnoteFile{YorkeBodenheimerLaughlin1993}

\bibitem[{{Yorke} et~al.(1995){Yorke}, {Bodenheimer}, and
  {Laughlin}}]{YorkeBodenheimerLaughlin1995}
{Yorke}, H.~W., {Bodenheimer}, P., and {Laughlin}, G. (1995).
\newblock {The formation of protostellar disks. 2: Disks around
  intermediate-mass stars}.
\newblock \emph{\apj} 443, 199--208.
\newblock \doi{10.1086/175514}
\bibAnnoteFile{YorkeBodenheimerLaughlin1995}

\bibitem[{{Zhao} et~al.(2018){Zhao}, {Caselli}, and {Li}}]{ZhaoCaselliLi2018}
{Zhao}, B., {Caselli}, P., and {Li}, Z.-Y. (2018).
\newblock {Effect of grain size on differential desorption of volatile species
  and on non-ideal MHD diffusivity}.
\newblock \emph{\mnras} 478, 2723--2736.
\newblock \doi{10.1093/mnras/sty1165}
\bibAnnoteFile{ZhaoCaselliLi2018}

\bibitem[{{Zhao} et~al.(2016){Zhao}, {Caselli}, {Li}, {Krasnopolsky}, {Shang},
  and {Nakamura}}]{Zhao+2016}
{Zhao}, B., {Caselli}, P., {Li}, Z.-Y., {Krasnopolsky}, R., {Shang}, H., and
  {Nakamura}, F. (2016).
\newblock {Protostellar disc formation enabled by removal of small dust
  grains}.
\newblock \emph{\mnras} 460, 2050--2076.
\newblock \doi{10.1093/mnras/stw1124}
\bibAnnoteFile{Zhao+2016}

\bibitem[{{Zhao} et~al.(2011){Zhao}, {Li}, {Nakamura}, {Krasnopolsky}, and
  {Shang}}]{Zhao+2011}
{Zhao}, B., {Li}, Z.-Y., {Nakamura}, F., {Krasnopolsky}, R., and {Shang}, H.
  (2011).
\newblock {Magnetic Flux Expulsion in Star Formation}.
\newblock \emph{\apj} 742, 10.
\newblock \doi{10.1088/0004-637X/742/1/10}
\bibAnnoteFile{Zhao+2011}

\end{thebibliography}
\end{document}